\documentstyle[preprint,prb,aps]{revtex}

\def \N {n^{-1}}

\def \n {n^{-1/2}}

\def \l {E}
\def \H {H}

\def \li {\lim_{n\to\infty}}

\def \T  {\hbox{tr~}}

\def \E {~{\bf E}\,}

\def \to {\rightarrow}

\def \C {{\bf C}_\pm }
\def \P {\Phi^{( n)}}

\def \a {\alpha^{(n)}}
\def \b {\beta^{(n)}}
\def \Ps {\Psi^{(n)}}

\def \Rb {{\bf R} }
\def \Cb {{\bf C} }

\def \ga {\gamma^{(n)}}
\def \ve {\varepsilon}

\begin{document}

\draft

\tighten

\title{ON ASYMPTOTIC PROPERTIES OF LARGE RANDOM MATRICES WITH INDEPENDENT ENTRIES}

\author{Alexei M. Khorunzhy}
\address{
B.I. Verkin Institute for Low Temperature Physics, Kharkov, 310164, Ukraine
        }
\author{Boris A. Khoruzhenko\cite{byline}}
\address{
Queen Mary \& Westfield College, University of London, London, E1 4NS, U.K.
        }
\author{Leonid A. Pastur}
\address{B.I. Verkin Institute for Low Temperature Physics, Kharkov, 310164,
Ukraine \\ and \\
Universit\'e Paris 7/ Denis Diderot, UMR 9994, Paris, 75251, France
       }

%\date{Version of \today}

\maketitle

\begin{abstract}
We study the normalized trace $g_n(z)=n^{-1} \mbox{tr} \,
(H-zI)^{-1}$ of the resolvent of $n\times n$ real symmetric
matrices $H=\big[(1+\delta_{jk})W_{jk}/\sqrt n\big]_{j,k=1}^n$  assuming that
their entries are independent
but not necessarily identically distributed random variables. We develop a
rigorous method of asymptotic
analysis of moments of $g_n(z)$ for $|\Im z| \ge \eta_0$ where  $\eta_0$ is
determined by
the second moment of $W_{jk}$. By using
this method we find the
asymptotic form of the expectation ${\bf E}\{g_n(z)\}$ and of the
connected correlator  ${\bf E}\{g_n(z_1)g_n(z_2)\}-{\bf E}\{g_n(z_1)\}
{\bf E}\{g_n(z_2)\}$. We also prove that the
centralized trace  $ng_n(z)- {\bf E}\{ng_n(z)\}$ has the Gaussian distribution
in the limit $n=\infty $. Basing on these results we present heuristic
arguments supporting the universality property of the local eigenvalue
statistics for this class of random matrix ensembles.

\end{abstract}


\section{Introduction}
\label{introduction}

Since the pioneer works of Wigner and  Dyson random matrix theory
(RMT) has been successfully used to describe the energy levels of complex
quantum systems: heavy nuclei, quantum chaotic systems, mesoscopic samples,
etc (see e.~g. Refs.\ \onlinecite{Mehta,Brody,Pichard,Bohigas,MF_1}).
Another rather broad field of the RMT applications is related to quantum field
theory: large colour limit of QCD, 2D quantum gravity and bosonic strings
(see e.g. Refs.  \onlinecite{BIZ,Demetrefi,Ginsparg}).

The phenomenological  nature of
the RMT approach that may be regarded as its certain drawback on the
one hand,
provides, on the other hand,
the model independent frameworks, that make the approach applicable to
a wide
variety of systems having different microscopic natures
and origins. These frameworks  assume
certain amount of
``robustness'' of the RMT models and results. In other words, it is believed
that ``sufficiently large'' number
of them should have no dependence or a rather weak one on
a random matrix ensemble used. This belief explains partly the
fact that  majority of the RMT ideas and applications  are based on results
obtained for the archetype Gaussian Ensembles (GE's)
and the Circular Ensembles (CE's). On the other hand, this belief requires
certain justification,
in particular extending the results known for the GE's and for the CE's to
other classes of ensembles.

The most referred to are the Gaussian Orthogonal Ensemble (GOE) of
random $n\times  n$ symmetric
matrices and the Gaussian Unitary Ensemble (GUE) of random $n\times  n$
Hermitian matrices. The density of the probability distribution in these
ensembles
has the form
\begin{eqnarray}
P_n(H) = Z_n^{-1} \exp[ -n\T  F(H)],
\label{P_n(H)}
\end{eqnarray}
where $F(x) = x^2/4w^2$ and $Z_n$ is the normalization constant.

The probability distribution (\ref{P_n(H)})  possesses two
important properties: i) it is invariant with respect to
either orthogonal or unitary transformations of $\Rb^n$ or  $\Cb^n$,
respectively; and ii) the matrix elements are independent random variables
(modulo the obvious symmetry conditions).

These properties of the GE's determine them uniquely and motivate
two classes of generalisations of the GE's.

The first class consists of ensembles having
an orthogonal or unitary invariant but not necessarily
matrix-element-independent
probability distribution. The typical
representatives are the ensembles with the probability distribution of the form
(\ref{P_n(H)})
in which $F(x)$ is an arbitrary bounded below and growing fast enough on
infinity function.
These  invariant ensembles can be used to describe physical
systems having
no preferential basis. They arose also in studying the large-$n$
limit in quantum field theory \cite{BIZ,Demetrefi,Ginsparg} and later found
other applications
\cite{Pichard,Ci,GP}.

Random matrices with invariant distributions show remarkable ``robustness''
(known as the universality) of spectral properties in the microscopic regime.
In this regime one scales
the energy so that the mean  distance  between nearest eigenvalues
remains of order unity as the dimension of matrices increases \cite{Mehta,Dy}.
Thus  one is able to study properties of a finite number of eigenvalues. The
universality of the level spacing distribution and other
microscopic (local) spectral characteristics
has been extensively discussed in recent theoretical physics and mathematical
literature. We refer the reader to a number of publications: Refs.
\onlinecite{Dy,KPW,P,BZ,W,PS}.

The second class consists of ensembles whose matrix elements  in a certain
basis are independent random variables, i.~e. the ensemble probability
distribution  factorizes into a product of distributions of the matrix elements
in this basis. The corresponding random matrices can be associated with
physical systems having a preferential basis and appear, in particular,  in
condensed matter physics
and theory of disordered systems.
This second class goes back to Wigner \cite{Wigner} and we shall refer to the
corresponding ensembles as Wigner ensembles (or Wigner matrices).

The subject of the present paper is the Wigner ensemble of $n \times n$ real
symmetric matrices of the form
\begin{eqnarray}
H=\left[ H_{jk}\right]_{j,k=1}^n, \;\;\;\; H_{jk}=(1+\delta_{jk} )W_{jk}/\sqrt
n,
\label{H2W}
\end{eqnarray}
\noindent where  $W_{jk}, j\le k$ are independent random variables such that
\begin{eqnarray}
{\bf E}\{W_{jk}\}=0, \;\;\;\; {\bf E}\{W_{jk}^2\}=w^2.
\label{W}
\end{eqnarray}
Here and thereafter ${\bf E}\{\cdot\}$ denotes averaging over all $W_{jk}, j\le
k$.

The distributions of $W_{jk}$'s may depend on  $(j,k)$, but we assume that they
are independent of $n$.
We make the latter assumption mainly  for the sake of technical simplicity. On
the other hand, this assumption allows one to consider all $W_{jk}$ on the same
probability space and to find an optimal form of a number of important facts
related to the Wigner ensembles (for example, the convergence with probability
1 in formulae (\ref{N2N_sc}) and (\ref{g2r}) below). If  $W_{jk}$'s are
independent Gaussian random variables, then the ensemble (\ref{H2W})-(\ref{W})
coincides with the GOE. This justifies the presence of the term with
$\delta_{jk}$ in (\ref{H2W}).

Macroscopic properties of Wigner ensembles are more or less well
understood. We call macroscopic the asymptotic regime in which the number of
eigenvalues in unit energy interval is proportional to $n$.
Discussing macroscopic properties of random matrices we have to mention first
of all the density of states (DOS) which is the simplest macroscopic
characteristic of the ensemble eigenvalue statistics. It turns out that under
rather natural and mild conditions on the distributions of $W_{jk}$ the DOS in
the Wigner ensemble  (\ref{H2W})-(\ref{W}) does not depend on the form of the
distributions of $W_{jk}$. This DOS is known as the Wigner semi-circle law (see
Eq.
(\ref {rho_semi}) below).
Other macroscopic spectral quantities such as the conductivity and the
interband light absorption coefficient  show the same ``robustness''
\cite{CMP,KKP,KKP_1}. Definition of these quantities requires some care for
Wigner matrices. However, the conductivity and the interband light absorption
coefficient can be defined and computed for the so-called band random matrices
and random operators  with independent matrix elements that are quite close to
the Wigner ensemble (\ref{H2W})-(\ref{W}) in their macroscopic properties both
technically and by results ( see e.g. Ref.\ \onlinecite{KKPS}).     As for the
microscopic scale, supersymmetry calculations \cite{MF_2} suggest
``robustness'' (universality) of spectral properties of the Wigner ensemble
(\ref{H2W})-(\ref{W})  as well  but evidence of this has not been rigorously
established so far.

Introduce the normalized eigenvalue counting function
\begin{eqnarray}
N_n(E)=n^{-1}\#\{E_j:  E_j \;\; \mbox{is an eigenvalue of} \;\; H \;\;
\mbox{and } \;\; E_j \le E\}.
\label{necf}
\end{eqnarray}
Wigner in the end of fifties proved \cite{Wigner} that in the case of
identically distributed  $W_{jk}$ having all moments  $N_n(E)$ converges in
probability as
$n\to \infty $ to a non-decreasing function $N_{sc}(E)$ (the semicircle law)
whose derivative (DOS) is
\begin{eqnarray}
\rho(E)=\left\{
\begin{array}{l l}
\begin{displaystyle} \frac{1}{2\pi w^2}\sqrt{4w^2-E^2} \end{displaystyle}
   & |E| \le 2w \\
 0 & |E| > 2w .
\end{array}
\right.
\label{rho_semi}
\end{eqnarray}

The modern formulation of Wigner's result is as
follows. Let us consider random matrices (\ref{H2W})-(\ref{W}) with mutually
independent arbitrary distributed entries defined on a common probability
space. Then the condition (the matrix analogue of the Lindeberg condition
of probability theory)
\begin{eqnarray}
\lim_{n \rightarrow \infty} \frac{1}{n^2}\sum_{j\le k}\; \int_{|x|>\nu n^{1/2}}
\;
x^2 d\mbox{Prob}[W_{jk} \le x] =0, \;\; \mbox{for any} \;\; \nu >0,
\label{Lin}
\end{eqnarray}
is sufficient   \cite{P_TMP} and necessary \cite{Girko} for the following
limiting relation
\begin{eqnarray}
\lim_{n \to \infty} N_n(E ) = N_{sc}(E).
\label{N2N_sc}
\end{eqnarray}
to hold for every $E$ with probability 1 \cite{ftnote}.
If we will not assume that $W{jk}$ are defined on the same probability
space or if their probability
distributions depend on $n$, then the same condition (\ref{Lin}) will
imply the convergence in probability
in (\ref{N2N_sc})

As usual in spectral theory, this result admits a natural reformulation in
terms of the resolvent (Green's function). Indeed, the normalized trace of the
resolvent
\begin{eqnarray}
g_n(z)=n^{-1}\T (H-zI)^{-1}
\label{g}
\end{eqnarray}
is simply the Stieltjes transform of $N_n(E)$:
\begin{eqnarray}
g_n(z) = {1\over n} \sum_{j=1}^n {1\over E_j  - z} =
 \int {dN_n(E) \over E- z}.
\label{ggg}
\end{eqnarray}

Denote the Stieltjes transform of the Wigner law (\ref{rho_semi}) by $r(z)$,
\begin{eqnarray}
r(z) = \int \frac{N_{sc}(dE)}{E - z}= \frac{-z+\sqrt{z^2-4w^2}}{2w^2}.
\label{r}
\end{eqnarray}
The obvious condition $\Im r(z) \Im z \ge 0$ determines the branch of the
square root in (\ref{r}).
Due to one-to-one correspondence between non-decreasing functions
and their Stieltjes transforms \cite{RS} (\ref{N2N_sc}) is equivalent to
the following limiting relation
\begin{eqnarray}
\li g_n(z) = r(z),
\label{g2r}
\end{eqnarray}
which  holds with probability 1 for any non-real $z$.

The relation (\ref {g2r}) and the obvious bound
$\vert g_n(z)\vert \leq {\vert \Im z\vert}^{-1}$
imply that the variance of $g_n(z)$ vanishes as $n \to \infty$ and hence the
moments
\begin{eqnarray}
m^{(p)}_n(z_1, \ldots ,z_p)={\bf E}\Big\{\prod_{l=1}^p g_n(z_l)\Big\}
\label{m}
\end{eqnarray}
factorize:
\begin{eqnarray}
m^{(p)}_n(z_1, \ldots  ,z_p)=\prod_{l=1}^p m^{(1)}_n(z_l)
 + o(1),\; n\to \infty
\label{fac}
\end{eqnarray}
This factorization, that follows already from the convergence in probability in
 (\ref {N2N_sc}) or in (\ref {g2r}) ,
is typical for the macroscopic regime and can be found hidden behind many
calculations in this regime. It is known in fact
since papers  Refs. \onlinecite {Wigner}, \onlinecite{MP}, \onlinecite{P_TMP}
and \onlinecite{Be}.

Since according to (11) $m^{(1)}_n(z)={\bf E}\{g_n(z)\}=r(z) + o(1)$, the
leading term of
$m^{(p)}_n(z_1, \ldots  ,z_p)$  is $\prod_{l=1}^p r(z_l)$.
It seems interesting from a number of points of view to find also  sub-leading
terms and  their dependence on the probability distributions of matrix
elements.
For instance these sub-leading terms are important  when we would like to go
beyond the macroscopic regime, when we are computing connected correlators of
$g_n(z)$, etc.

For the Gaussian entries respective corrections were
studied in Ref.\ \onlinecite{VWZ}
where  the formal perturbation theory with respect to $H_{jk}$
and the respective diagrammatic technique were applied. This approach is an
adaptation
of the technique developed in Ref.\ \onlinecite{OW} in
order to construct the $1/n$ expansion for  the random operator describing
disordered systems on ${\bf Z}^d$ with
$n$ orbitals per site.

In Ref.\ \onlinecite{KhKP} we suggested
an approach that allows for the rigorous treatment of this problem in the
general case of independent and arbitrary not necessarily identically
distributed
matrix elements.
Our approach allows us to estimate remainders in respective asymptotic formulae
and we show that these estimates are in a sense optimal. The approach is also
free to the large extent from the  cumbersome combinatorial
problem of rearranging diagrams which is neccessary in order to carry out
various ``dressing'' procedures. In particular,
the dressing  procedure that replaces the "bare" Green function $-1/z$ by
$\lim_{n \to \infty} \; {\bf E}\{g_n(z)\} $ is automatic in our approach.
Following Ref.\ \onlinecite{KhKP} one is able to find as many terms in the
asymptotic  expansion of $m^{(p)}_n(z_1, \ldots  ,z_p)$ as needed, though the
technical difficulties increase with the order.

In the present  paper we use the general scheme of Ref.\ \onlinecite{KhKP} in
order to compute first terms in the
asymptotic expansion of ${\bf E}\{g_n(z)\}$. We also prove that if the
distributions of $W_{jk}$ satisfy the Lindeberg
condition (\ref{Lin}) with $x^2$ in the integral replaced by $x^4$,  and if in
addition to (\ref{W}) the fourth moments
of $W_{jk}$ do not depend on  $(j,k)$, then
\begin{eqnarray}
F_n(z_1, z_2)=
m^{(2)}_n(z_1,z_2)-m^{(1)}_n(z_1)m^{(1)}_n(z_2)=n^{-2}f(z_1,z_2)+o(n^{-2}),
\label{cov}
\end{eqnarray}
\noindent where
\begin{eqnarray}
\lefteqn{
f(z_1,z_2)=} \label{f} \\
 & & \frac{2w^2}{[1-w^2r^2(z_1)][1-w^2r^2(z_2)]}
\left[\frac{r(z_1)-r(z_2)}{z_1-z_2}\right]^2 +
\frac{2\sigma r^3(z_1)r^3(z_2)}{[1-w^2r^2(z_1)][1-w^2r^2(z_2)]}. \nonumber
\end{eqnarray}
and  $\sigma={\bf E}\{W_{jk}^4\}-3{\bf E}^2\{W_{jk}^2\}$ is the excess of
$W_{jk}$. We  also establish a limit theorem for
the centralized trace $ng_n(z) - {\bf E}\{ng_n (z)\}$  of the resolvent.

Unfortunately, our approach gives a bound for the remainder term in (\ref{cov})
containing a power of $|\Im z_1 \Im z_2|^{-1}$ as a factor. Thus we cannot
treat rigorously the microscopic regime which requires $\Im z \propto 1/n$.
On the other hand, the first  term (15) of the asymptotic formula (14) is well
defined in this regime and coincides with respective exact expression known for
the GOE, provided that the latter is considered for large level spacings and
is smoothed over an interval $\Delta$ such that $1/n \ll |\Delta| \ll 1$ in
proper units (see Section VI). We feel therefore, that by using our procedure
of the computing of
 corrections, i.e.
keeping the imaginary part of energy fixed when $n$ goes to infinity and then
letting $\Im z$ go to zero , one may treat energy intervals that are very large
on the microscopic scale. On this intermediate scale the second term in the
r.h.s. of (\ref{f}) which contains the probability distribution excess $\sigma
$ vanishes and the above mentioned universality restores.

Our article is organized as follows. In section  \ref{section2} we present our
basic tools. In sections \ref{section3}  we  calculate first terms of the
asymptotic expansion for ${\bf E}\{g_n (z)\}$. In section  \ref{section4} we
give a simple proof of (\ref{cov})-(\ref{f}) with $o(n^{-2})$ replaced by
$O(n^{-5/2})$ provided that 5th absolute moment of $W_{jk}$ is uniformly
bounded. This result was cited without proof in Ref.\  \onlinecite{KhKP}. In
section \ref{section5}  we treat the general case of $W_{jk}$ satisfying the
higher order Lindeberg condition mentioned above. We prove that the
fluctuations of $ng_n(z)$ around its mean value become Gaussian in
 the limit $n \rightarrow \infty$ and that the  covariance of the limiting
Gaussian function is $f(z_1,z_2)$, thus proving
(\ref{cov})-(\ref{f}) in the general case.  Section \ref{section6}  contains
 a discussion of some implications of our results.

%%%%%%%%%%%%%%%%%%%%%%%%%%%%   sections 2   %%%%%%%%%%%%%%%%%%%%%%%%%%
\section{Preliminaries}
\label{section2}
In this section we present our basic technical tools.

\begin{itemize}

\item[(i)] If $\xi$ is a real-valued random variable such that
${\bf E} \{|\xi|^{p+2}\}<\infty$ and if $f(t)$ is a complex-valued function of
a real variable such that its first $p+1$ derivatives are continuous and
bounded, then
\begin{eqnarray}
{\bf E}\{\xi f(\xi)\}=\sum_{a=0}^{p} \frac{\kappa_{a+1}}{a!} {\bf E}
\{f^{(a)}(\xi)\} + \varepsilon,
\label{im_nf}
\end{eqnarray}
\noindent where $\kappa_a$ are the semi-invariants (cumulants)  of $\xi$,
$|\varepsilon| \le C\sup_t |f^{(p+1)}(t)| \, {\bf E}\{|\xi |^{p+2}\}$ and the
quantity $C$ depends on $p$ only.

The semi-invariants can be expressed in terms of the moments. If ${\bf E}
\{\xi\} =0$ (the case we shall deal with) and $\mu_a = {\bf E}\{\xi^a\}$, then
few first such relations are: $\kappa_1=\mu_1=0$, $\kappa_2=\mu_2$,
$\kappa_3=\mu_3$, $\kappa_4=\mu_4-3\mu_2^2$, $\kappa_5=\mu_5-10\mu_3 \mu_2$,
$\kappa_6=\mu_6-15\mu_4 \mu_2 -10\mu_3^2 -30\mu_2^3$, etc. For a Gaussian
random  variable with zero mean, all semi-invariants but
$\kappa_2$ vanish and (\ref{im_nf}) reduces to the exact relation
\begin{eqnarray}
{\bf E}\{\xi f(\xi)\}={\bf E}\{\xi^2\} {\bf E} \{f^{\prime} (\xi)\},
\label{nf}
\end{eqnarray}
\noindent which can directly be checked integrating  the l.h.s. of (\ref{nf})
by parts. This is only the case , when formula (\ref{im_nf})
contains finite
number of terms for non-polynomial $f$'s. Indeed, according to the
Marcynkiewicz theorem \cite{Lu}
if all but finite number of cumulants are zero, then only first and second
can be nonzero.

\item[(ii)] For any matrix $A=[A_{\alpha \beta}]_{\alpha, \beta =1}^n$
\begin{eqnarray*}
\frac{\partial }{\partial A_{jk}} \left(A^{-1}\right)_{\alpha \beta} = -
\left(A^{-1}\right)_{\alpha j}\left(A^{-1}\right)_{k \beta}
\end{eqnarray*}
provided $A^{-1}$ exists.
For the resolvent $G$ of a real symmetric matrix $H$ this becomes
\begin{eqnarray}
\frac{\partial G_{\alpha \beta}}{\partial H_{jk}} =
\left\{
\begin{array}{l l l}
-G_{\alpha j} G_{k \beta}, & & j=k \\
-G_{\alpha j}G_{k \beta} -
G_{\alpha k} G_{j \beta}, & & j\not= k\\
\end{array}
\right.
\label{dG}
\end{eqnarray}

\item[(iii)] For any two real symmetric matrices and any non-real $z$ the
resolvent identity
\begin{eqnarray}
(H_2-zI)^{-1}=(H_1-zI)^{-1}-(H_1-zI)^{-1}(H_2-H_1)(H_2-zI)^{-1}
\label{res_id}
\end{eqnarray}
is valid. In particular, if $H_2=H$, $H_1=0$ and $G=(H-zI)^{-1}$, then
\begin{eqnarray}
G_{jm}=z^{-1}\delta_{jm}+z^{-1}\sum_{k=1}^n G_{jk}H_{km}
\label{res}
\end{eqnarray}
\end{itemize}

Let $H$ belong to the Wigner ensemble (\ref{H2W})-(\ref{W}).
For a fixed complex $z$ consider complex-valued random variable
$g_n(z)=n^{-1} \T (H-zI)^{-1}$. Define its variance as
\begin{eqnarray}
{\bf E} \{|g_n(z)- {\bf E} \{g_n(z) \}|^2  \}=F_n(z, z^\dagger).
\label{F}
\end{eqnarray}
and define also the domain in the complex plane as follows
\begin{eqnarray}
U_0 = \{z \in \C: \vert \Im z \vert \ge 2w\}.
\label{U_0}
\end{eqnarray}

We use $^\dagger$ to denote complex conjugate and the sub-index $C$ to
indicate centering to zero mean. For instance, $g^C_n(z)=g_n(z)- {\bf E}
\{g_n(z)$ and thus we can rewrite  (\ref{F})  as
\begin{eqnarray}
F_n(z, z^\dagger)=
{\bf E} \{|g^C_n(z)|^2  \}={\bf E} \{g_n(z)g^C_n(z^\dagger)  \}.
\label{VaR}
\end{eqnarray}
We will write $O(n^{-p})$ in asymptotic formulae for the remainders
having an uniform (with respect
$z \in U_0$) upper bound of the form $Cn^{-p}$ where $C$ does not depend on
$n$.
In fact the bounds we are able to derive contain $1/(1-|\Im z|^2 /2w^2)$
(see e.g. formula (29) below for the simplest case). Thus  $C$ is finite for
any fixed $z$ satisfying
$|\Im z| > \sqrt 2 w$. But we prefer to use $|\Im z| \ge 2 w$ in favour of
uniformity of the bounds with respect to $z \in U_0$.

\begin{itemize}

\item[(iv)] Let $H$ belong to the Wigner ensemble (\ref{H2W})-(\ref{W}).
Assume that the 5th absolute moment of the random variables $W_{jk}$ is
uniformly bounded, i.~e. $\sup_{j\le k} {\bf E} \{|W_{jk}|^5\} < +\infty$, and
that
$z\in U_0$.
Then
\begin{eqnarray}
{\bf E} \{|g_n(z)- {\bf E} \{g_n(z) \}|^2  \}={\bf E} \{|g^C_n(z)|^2  \}=
O(n^{-2}), \;\; {\mbox{as}} \;\; n\to \infty
\label{var_g}
\end{eqnarray}

\end{itemize}
\vskip 0.5cm

Let us comment on (i)-(iv). Facts (ii) and (iii) are well known. The
``decoupling'' formula (\ref{im_nf}) is simple to understand in the case when
$\xi $ has all moments and $f(x)$ belongs to the Schwartz space. Indeed, by
using the Parseval relation for the Fourier transforms we can rewrite the
l.h.s. of (\ref{im_nf}) as
\begin{eqnarray}
{i \over 2\pi}
\int_{-\infty }^{\infty }{d \over d t} F^\dagger (t)\Pi (t)d t=
-{i \over 2\pi} \int_{-\infty }^{\infty }F^\dagger (t){d \over d t} \Pi (t)d t
\label{pars}
\end{eqnarray}
where
\begin{eqnarray*}
F(t)=\int_{-\infty }^{\infty }e^{i \xi t}f(\xi )d \xi
\;\;
\mbox{and} \;\;
\Pi (t)=\int_{-\infty }^{\infty }e^{i \xi t}d P(\xi )
\end{eqnarray*}
are the Fourier transforms of $f(\xi )$ and of the probability distribution
$P(\cdot )$ of $\xi$, respectively. Now, if we take into account that
\begin{eqnarray*}
\Pi (t)=\sum_{a=0}^{\infty } \frac{{(i t)}^{a}\mu _{a}}{a!} \;\;
\mbox{and} \;\;
u(t)\equiv \log \Pi (t)=\sum_{a=1}^{\infty } \frac{{(i t)}^{a}\kappa _{a}}{a!}
\end{eqnarray*}
we can rewrite the r.h.s. of (\ref{pars}) as
\begin{eqnarray*}
-{i \over 2\pi}\int_{-\infty }^{\infty }F^\dagger (t)u'(t)e^{u(t)}d t &=&
-{i \over 2\pi }\sum_{a=0}^{\infty } \frac{\kappa _{a+1}}{a!}\int_{-\infty
}^{\infty }{(i t)}^{a+1} F^\dagger (t) \Pi (t) d t \\
 &=&
\sum_{a=0}^{\infty } \frac{\kappa_{a+1}}{a!} {\bf E} \{f^{(a)}(\xi)\}
\end{eqnarray*}
where we again used the Parseval relation. The latter formula is obviously
(\ref{im_nf}) for $p=\infty $. The case when $\xi$ has a finite number of
moments and $f(\xi)$ has respective number of derivatives requires certain
technicalities which we will not discuss here.

The bound (iv) plays an important role in many questions of random matrix
theory and its applications. In fact, it is the simplest of bounds for
connected correlators (cumulants) of $g_n(z)$ or, more generally, for cumulants
of linear statistics of the eigenvalues ( i.e. sums
$n^{-1}\sum_{j=0}^{n}\phi(E_j)$
where $\phi(E)$ is smooth enough). We are going to present a detailed
derivation of these bounds and asymptotics (both, for the Wigner ensembles and
unitary invariant ensembles) in a subsequent publication.

Here we only outline the scheme of the derivation of the bound (iv) considering
the simplest case of the GOE
and treating it as a representative of the Wigner ensembles, i.e. ensembles
with independent entries. Set $r_n(z)={\bf E} \{g_n(z) \}$.
Then, according to  (\ref{nf}), (\ref{dG}) and (\ref{res}) we have the relation
\begin{eqnarray}
r_n(z)=-{1 \over z}-{w^2 \over z} {\bf E} \{g^2_n(z)\}+
{w^2 \over n^2z} {\bf E} \{\T G^2\}
\label{id_1}
\end{eqnarray}
Applying similar arguments to ${\bf E} \{|g^2_n(z)|\}$  and using  (\ref{id_1})
we obtain the analogous relation for the variance (\ref{VaR})
\begin{eqnarray}
F_n= - {w^2 \over z}E\{ g_n^2(z)
g_n^C (z^\dagger) \}-
{w^2 \over n^2z}{\bf E}\{[g_n^C (z)]^\dagger \T G^2\}
-{2w^2 \over n^3z}{\bf E}\{\T G(G^*)^2\}
\label{id_2}
\end{eqnarray}
where $G^*=(H-z^\dagger I)^{-1}$.
By using  the identity  $E\{ g_n^2(z) g_n^C (z^\dagger ) \}=E\{( g_n(z) +
E\{ g_n(z)\} )
|g_n^C (z)|^2 \}$, Cauchy-Schwarz inequality and the inequality
(\ref{Gn_b}) below we can show that the first term in the r.h.s
of (\ref{id_2}) is bounded above by $2w^2\eta ^{-2}F_n$,
where $\eta =|\Im z|$, the second
is bounded by
$w^2(n\eta ^2)^{-1}F_n^{1/2}$ and the third
is bounded by $2w^2(n^2\eta ^4)^{-1}$. As the result we obtain the following
inequality for
$\eta^2>2w^2$
\begin{eqnarray}
(1- {2w^2 \over \eta ^2})F_n-
{w^2 \over n\eta ^2}F_n^{1/2} - {2w^2 \over n^2\eta ^4} \le 0.
\label{qe}
\end{eqnarray}
which implies  that
\begin{eqnarray}
F_n={\bf E}\{ |g_n^C(z)|^2 \} \equiv {\bf E}
\{ |g_n(z)- {\bf E} \{g_n(z)\}|^2\}
 \le {C_1\over n^{2}}
\label{gb}
\end{eqnarray}
where $ C_1=\eta^{-2}(\epsilon - 1)^{-2}C_1(\epsilon)$, $\epsilon =
\eta {^2}/2w^2,\, \, \, 1<\epsilon <\infty$ and $C_1(\epsilon)$
is finite for $1\le \epsilon <\infty$
Thus we have obtained (\ref{var_g}) for the GOE. This is a simplest but typical
bound that can be obtained by our method. In the general case of non-Gaussian
$W_{jk}$'s one has to iterate the resolvent identity (20) and use (16) instead
of (17),
truncating this procedure on the proper step and estimating remainders by
variants of arguments presented above.

The bound (\ref{gb}) and (\ref{id_1})allows us to
prove (\ref{N2N_sc}) and  (\ref{g2r}) for the GOE.  Indeed, combining
(\ref{id_1}) and  (\ref{gb}) we obtain
\begin{eqnarray*}
|r_n(z)+{1 \over z}+{w^2 \over z}r^2_n(z)|
\le {C_{2}\over n}
\end{eqnarray*}
where $ C_2$ has same properties as  $ C_1$ in  (\ref{gb}).
This bound and standard compactness arguments show that  any limit point
$r(z)$ of the sequence $\{r_n(z)\}$ satisfies the equation
\begin{eqnarray}
w^2r^2(z)+zr(z)+1=0.
\label{quadra}
\end{eqnarray}
for $ |\Im z| \ge \eta_0  > 0.
$
Since this  equation
has the unique solution  (\ref{r}) satisfying $\Im r(z) \Im z \ge 0$, we
conclude that uniformly
in $|\Im z| \ge \eta_0  > 0 \,\,\,
\lim_{n \to \infty} r_n(z) = r(z)$ where $r(z)$ is given by (\ref{r}). Besides,
since  the Gaussian $W_{jk}$ satisfying (\ref{W})
can be defined on the same  probability
space we conclude from (\ref{gb}) and the  Borel-Cantelli lemma that
(\ref{g2r}) and (\ref{N2N_sc}) and  are valid.

%%%%%%%%%%%%%%%%%%%%%%% section 3 %%%%%%%%%%%%%%%%%%%%%%%%%%%%%%%%%

\section{
asymptotic expansion for
${\bf E}\lowercase{
                    \{ g_n(z) \}$
                  }
        }

\label{section3}

We recall our notation $m_n^{(1)}(z)$ for the mean value of $g_n(z)= n^{-1}\T
(H-zI)^{-1}$. In this section we  prove the following
\vskip 0.5cm

{\bf Theorem 1}
{\sl
Consider the Wigner ensemble of random  real symmetric
matrices with independent entries defined by  (\ref{H2W})-(\ref{W}).  Assume
additionally that the
third and fourth moments of $W_{jk}$ do not depend on
$j$ and $k$ and that  $\widehat{\mu_5}
=\sup_{j\le k} {\bf E} \{|W_{jk}|^5\} < +\infty$.

Then the following  asymptotic formula
\begin{eqnarray}
m_n^{(1)}(z)=r(z)
\left\{
1+ \frac{1}{n} \left[
\frac{w^2r^2(z)}{[1-w^2r^2(z)]^2}
 +  \frac{\sigma r^4(z)}{1-w^2r^2(z)}\right]
\right\}
+O(n^{-3/2}).
\label{n2lo}
\end{eqnarray}
holds for any $z \in U_0$ ($U_0$ is defined in (\ref{U_0}))}

\vskip 1cm {\sl Proof.} By the resolvent identity (\ref{res_id})-(\ref{res}),
\begin{eqnarray}
m^{(1)}_n(z)=-z^{-1} + (zn)^{-1}\sum_{j,m=1}^n {\bf E} \{ G_{jm}H_{mj}\}.
\label{g_n}
\end{eqnarray}
If we were following the conventional perturbational-diagrammatic approach
trying to develop the asymptotic expansion for ${\bf E} \{ g_n(z)\}$, we would
repeatedly iterate the resolvent identity selecting on each step the terms that
contribute to the leading and sub-leading terms. The obvious drawback of such
approach is that infinitely many iterations are needed and in the non-Gaussian
case, when there is no analogue of the Wick theorem,  the diagrammatic approach
is rather complicated.

We propose making use of (\ref{im_nf}) instead of iterating the resolvent
identity. For each pair $(j,m)$, $G_{jm}$ is a smooth function of $H_{mj}$ and
its derivatives are bounded
because of (\ref{dG}) and the inequality
\begin{eqnarray}
|G_{jm}| \le \| G\| \le |\Im z|^{-1}
\label{Gn_b}
\end{eqnarray}
which holds for the resolvent of any real symmetric matrix. In particular,
$|D^4_{mj} G_{jm}| \le C |\Im z|^{-5}$
where  $C$ is an absolute constant. Here and thereafter  we use notation
$D_{mj}$ for $\partial /\partial H_{mj}$.

According to  (\ref{H2W})-(\ref{W}) and our assumptions,  the fifth absolute
moment of $H_{mj}$ is of order $n^{-5/2}$. Thus applying (\ref{im_nf}) (with
$p=3$) to each of the summands in the r.h.s. of (\ref{g_n}) one finds that
\begin{eqnarray}
zm^{(1)}_n(z)=-1 +
\sum_{a=1}^3 \frac{1}{n^{(a+3)/2}}\!\!
\sum_{j,m=1}^n \!
\frac{\kappa_{a+1} (1+\delta_{jm})^{(a+1)/2} }{a!}
{\bf E} \{D_{mj}^a G_{jm} \} + \varepsilon_n,
\label{next_g_n}
\end{eqnarray}
where $\kappa_a$ are the semi-invariants of $W_{mj}$ and
\begin{eqnarray*}
|\varepsilon_n| \le  \frac{C}{n^{3/2}}
\frac{\widehat{\mu_5}}{|\Im z|^5}.
\end{eqnarray*}

Obviously, $G$ is a complex symmetric matrix, i.~e. $G_{jm}=G_{mj}$. By
(\ref{dG}),
$D_{mm}^a G_{mm} = a! G_{mm}^{a+1}$ and
\begin{eqnarray}
-D^1_{mj} G_{jm} & = & G_{jm}^2 + G_{jj}G_{mm}  \label{D1}\\
 D^2_{mj} G_{jm} & = & 2 G_{jm}^3 + 6G_{jm}G_{jj}G_{mm} \label{D2}\\
-D^3_{mj} G_{jm} & = & 6 G_{jm}^4 + 36G_{jm}^2G_{jj}G_{mm} +6G_{jj}^2G_{mm}^2
\label{D3}
\end{eqnarray}
for distinct $j$ and $m$. Let us set $\kappa_2=w^2$ and $\kappa_4=\sigma$ in
(\ref{next_g_n}). Then, as a consequence
of (\ref{D1})-(\ref{D3}),
\begin{eqnarray}
zm^{(1)}_n(z)=-1-w^2 m^{(2)}_n(z,z) -n^{-1}\big[w^2{\bf E} \{ c_n(z)\} + \sigma
{\bf E}\{d_n^2(z)\}\big]  + \varepsilon_n,
\label{E_g_n}
\end{eqnarray}
where
\begin{eqnarray}
c_n(z)=\frac{1}{n}\sum_{j,m=1}^n G_{jm}^2, \;\;
d_n(z)=\frac{1}{n}\sum_{m=1}^n G_{mm}^2
\label{cd}
\end{eqnarray}
and
\begin{eqnarray}
|\varepsilon_n| \le \frac{C}{n^{3/2}}\left( \frac{|\kappa_3|}{|\Im z|^3} +
\frac{|\kappa_4|}{|\Im  z|^4} + \frac{\widehat{\mu_5}}{|\Im z|^5}\right).
\label{eps_n}
\end{eqnarray}
provided $|\Im z| " 2w$ and $n$ is large enough.

To infer (\ref{E_g_n})-(\ref{eps_n}) from  (\ref{next_g_n}),  notice first that
for the sum over $j=m$ in the r.h.s. of  (\ref{next_g_n}) we have the bound
\begin{eqnarray*}
 \frac{1}{n^{(a+3)/2}} \left| \sum_{m=1}^n D_{mm}^a G_{mm} \right|
\le \frac{C}{n^{(a+1)/2} |\Im z|^{a+1}}\propto
 \frac{1}{n^{(a+1)/2}} \le \frac{1}{n^{3/2}}, \;\; a=2,3.
\end{eqnarray*}
for all realizations of $W_{jk}$.
Therefore, being interested in the leading-order and $1/n$-order terms of
$m^{(1)}_n(z)$ we can omit $\delta_{jm}$ from the factor in front of the second
and third derivatives. As for the first derivatives, it follows from
(\ref{D1}),that  for all $j$ and $m$ $(1+\delta_{jm})
D^1_{jm} G_{jm} = G_{jm}^2+G_{jj}G_{mm}$  and  the term arising from
$\delta_{jm}$ contributes to $1/n$-order term in the asymptotic expansion of
$m^{(1)}_n(z)$.

Now, $G_{jm}^2$ in the r.h.s. of (\ref{D1}) makes ${\bf E} \{c_n(z)\}$ in
(\ref{E_g_n}) and $G_{jj}G_{mm}$ does
$m^{(2)}_n(z,z)$.The term containing $G_{jj}^2G_{mm}^2$ in the r.h.s. of
(\ref{D3}) leads to ${\bf E} \{ d^2_n(z) \}$ in (\ref{E_g_n}) and  the rest in
the r.h.s. of (\ref{D2}) and (\ref{D3})  contributes to $\varepsilon_n$ in
(\ref{E_g_n}). Corresponding bounds for  the terms coming from  $G_{jm}^3$ in
(\ref{D2}) and from  $G_{jm}^4$ and $G_{jm}^2 G_{jj}G_{mm}$ in (\ref{D3})
result from the simple inequality
\begin{eqnarray}
n^{-1} \sum_{j,m=1}^n |G_{jm}|^p \le |\Im z|^{-p}, \; \; p \ge 2
\label{p_bound}
\end{eqnarray}
which holds for the  resolvent of any real symmetric matrix. Estimating
the term coming from $G_{jm}G_{jj}G_{mm}$ in the r.h.s. of (\ref{D2}) requires
a longer calculation. Set
\begin{eqnarray}
h_n(z)=\frac{1}{n}\sum_{j,m=1}^n G_{jm}G_{jj}G_{mm}.
\label{fn}
\end{eqnarray}
Substitute the r.h.s. of (\ref{res}) for
$G_{jm}$ in $h_n(z)$. Then
\begin{eqnarray}
z{\bf E} \{ h_n(z)\}=-\frac{1}{n}\sum_{m=1}^n {\bf E} \{G_{mm}^2 \}
-w^2{\bf E} \{ g_n(z)h_n(z) \} + O( n^{-1/2} )
\label{f_n}
\end{eqnarray}
as it follows from (\ref{im_nf}), (\ref{D1})-(\ref{D3}) and simple resolvent
bounds like (\ref{Gn_b}) or (\ref{p_bound}).
Here and   below we use notation $O(n^{-p})$ for remainders admitting the upper
bound $Cn^{-p}$, where $C$ does not depend on $n$ for $|\Im z|\geq 2w$.

According to (24) the variance of $g_n(z)$ is of order $n^{-2}$ under our
assumptions. In other words
\begin{eqnarray}
m^{(2)}_n(z,z)=[m^{(1)}_n(z)]^2+O(n^{-2})
\label{m_2}
\end{eqnarray}
if $z \in U_0$. Obviously,
\begin{eqnarray*}
{\bf E} \{ h_n(z)g_n(z)\}-{\bf E} \{ h_n(z)\}{\bf E} \{ g_n(z)\}=
{\bf E} \{ h_n(z)[g_n(z)-{\bf E}\{g_n(z) \}]\}
\end{eqnarray*}
and by the Cauchy-Schwarz inequality
\begin{eqnarray*}
{\bf E} \{ h_n(z)g_n(z)\}={\bf E} \{ h_n(z)\}m^{(1)}_n(z) +O(n^{-1}).
\end{eqnarray*}
Therefore by (\ref{f_n}),
\begin{eqnarray}
\big[z-w^2m^{(1)}_n(z)\big] {\bf E} \{ h_n(z)\}= -\frac{1}{n}\sum_{m=1}^n {\bf
E} \left\{ G_{mm}^2 \right\} + O(n^{-1/2})
\label{E_f_n}
\end{eqnarray}
and ${\bf E} \{ h_n(z)\}$ is of order unity. The term we wish to estimate is
\begin{eqnarray*}
\frac{2\kappa_3}{n^{5/2}}\sum_{j,m=1}^n {\bf E} \{G_{jm}G_{jj}G_{mm}\}=
\frac{2\kappa_3}{n^{3/2}} {\bf E} \{h_n(z) \}.
\end{eqnarray*}
and from (\ref{E_f_n}) we see it is of order $n^{-3/2}$. This proves
(\ref{E_g_n})-(\ref{eps_n}).

The calculation above is typical of our approach and uses (\ref{im_nf}) and
(\ref{res}) combined with simple resolvent bounds on different stages. In what
follows we shall  (\ref{cov}) and (\ref{f})often use similar calculations
omitting details.

Equations (\ref{E_g_n})-(\ref{eps_n}) and (\ref{m_2}) imply that
\begin{eqnarray}
m^{(1)}_n(z)=r(z)+O(n^{-1}),
\label{lo_m}
\end{eqnarray}
where $r(z)$ solves the equation (\ref{quadra}).
Because of (\ref{ggg}) $m_n^{1}(z)$ as a function of $ z$ must satisfy the
inequality $\Im r(z)\Im z \geq 0$. This restriction fix the branch of the
square root in the expression for the solutions of (\ref{quadra}). Thus  $r(z)$
coincides with the  Stieltjes transform (\ref{r}) of the semi-circle law
(\ref{rho_semi}), as expected.

Once the leading term of $m^{(1)}_n(z)$ is found, we can  proceed with finding
the sub-leading term. From (\ref{E_g_n})
it is clear that performing this task requires calculating the leading-order
terms of  ${\bf E} \{ c_n(z)\}$ and ${\bf E} \{ d^2_n(z)\}$.  Substitute the
r.h.s. of (\ref{res}) for one of $G_{jm}$ in $c_n(z)$ and apply (\ref{im_nf}).
As a result,
\begin{eqnarray*}
z{\bf E} \{ c_n(z)\} =-m^{(1)}_n(z) - 2w^2 {\bf E} \{ c_n(z)g_n(z)\}
+O(n^{-1}).
\end{eqnarray*}
By (\ref{var_g}),
\begin{eqnarray}
z{\bf E} \{ c_n(z)\} =-m^{(1)}_n(z) - 2w^2 m^{(1)}_n(z){\bf E} \{ c_n(z)\}
+O(n^{-1})
\label{E_c_n}
\end{eqnarray}
and in the leading order
\begin{eqnarray*}
{\bf E} \{ c_n(z)\}=-m^{(1)}_n(z)\big[z+2w^2m^{(1)}_n(z)\big]^{-1}.
\end{eqnarray*}
Taking into account (\ref{lo_m})-(\ref{quadra}) we conclude that
\begin{eqnarray}
{\bf E} \{ c_n(z)\}=r^2(z)\big[1-w^2r^2(z)\big]^{-1}+O(n^{-1}).
\label{lo_c}
\end{eqnarray}

Now, calculate the leading-order term of ${\bf E} \{ d^2_n(z)\}$. Recall that
according to (\ref{var_g}) the variance of $g_n(z)$ is of order $n^{-2}$. By
(\ref{g2r}), $g_n(z)=n^{-1}\sum_{m=1}^n G_{mm}$
 converges almost surely to $r(z)$ as $n \to \infty$. Or, put it into another
way, the Cesaro limit of $G_{mm}$ is $r(z)$.
This suggests that the Cesaro limit of $G_{mm}^2$ should be equal to $r^2(z)$,
or in other words $d_n(z)$ should converge
almost surely to $r^2(z)$.  Therefore ${\bf E} \{ d^2_n(z)\}$ should  converge
to $r^2(z)$.

To prove the convergence rigorously and to estimate its rate, we first note
that the variance of $d_n(z)$ is of order
$n^{-1}$ if $z \in U_0$  (this can be proved following calculations of
Appendix \ref{appendixB}). Therefore
\begin{eqnarray}
{\bf E} \{ d^2_n(z)\}={\bf E} \{ d_n(z)\}^2 +O(n^{-2}).
\label{var_d}
\end{eqnarray}
Thus, it suffices to find the leading-order term of ${\bf E} \{ d_n(z)\}$.

Again, as in the case of $c_n(z)$,  substitute the r.h.s. of (\ref{res})
($j=m$) for one of $G_{mm}$ in $d_n(z)$ and apply (\ref{im_nf}). As a result,
\begin{eqnarray*}
z{\bf E} \{ d_n(z)\} =-m^{(1)}_n(z) - w^2 {\bf E} \{ d_n(z)g_n(z)\}
+O(n^{-1/2}).
\end{eqnarray*}
By (\ref{var_g}),
\begin{eqnarray*}
z{\bf E} \{ d_n(z)\} =-m^{(1)}_n(z) - w^2 m^{(1)}_n(z) {\bf E} \{ d_n(z)\}
+O(n^{-1/2})
\end{eqnarray*}
and
\begin{eqnarray*}
{\bf E} \{d_n(z)\} = -m^{(1)}_n(z) \big[ z+w^2m^{(1)}_n(z)
\big]^{-1}+O(n^{-1/2}).
\end{eqnarray*}
Finally by (\ref{lo_m}) and (\ref{quadra}),
\begin{eqnarray}
{\bf E} \{d_n(z)\} =r^2(z)+O(n^{-1/2})
\label{d_n_z}
\end{eqnarray}
and by (\ref{var_d}),
\begin{eqnarray}
{\bf E} \{d^2_n(z)\} =r^4(z)+O(n^{-1/2}).
\label{lo_d}
\end{eqnarray}

Now we are in a position  to find the sub-leading term of $m^{(1)}_n(z)$.
Collect (\ref{E_g_n}), (\ref{m_2}), (\ref{lo_c}) and (\ref{lo_d}) and write
\begin{eqnarray*}
zm^{(1)}_n(z)=-1-w^2 [m^{(1)}_n(z)]^2 -\frac{1}{n}
\left[
\frac{w^2r^2(z)}{1-w^2r^2(z)}
 + \sigma r^4(z)
\right]+O(n^{-3/2}).
\end{eqnarray*}
In view of (\ref{lo_m}) and (\ref{quadra}) this relation is obviously
equivalent to the statement of the theorem, i.e. to
the asymptotic formula (\ref{n2lo}). The theorem is proved.

\vskip 1cm {\bf Remarks}

1) Our bound for the remainder in (\ref{n2lo}) is an optimal one. For assuming
the 6th absolute moment of $W_{jk}$ to be uniformly bounded and keeping one
more term when applying
(\ref{im_nf}), we can find the term of order $n^{-3/2}$ in
the asymptotic expansion of $m_n^{(1)}(z)$. This term is
proportional to $\kappa_3={\bf E} \{ W_{jk}^3\}$

2) If the distributions of $W_{jk}$ are such that ${\bf E} \{ W^3_{jk} \}=0 $,
then  the bound  $O(n^{-3/2})$ for the remainder in
(\ref{n2lo}) can be strengthened to $O(n^{-2})$. For terms of order $n^{-3/2}$
appear in (\ref{n2lo}) due to contribution of $\kappa_3 n^{-3/2} {\bf E} \{
h_n(z) \}$ to  $\varepsilon_n$ in (\ref{E_g_n}) and also because of
(\ref{lo_d}). If
 ${\bf E} \{W^3_{jk} \}=0 $, then $\kappa_3=0$ and we can prove that the
remainder in (\ref{lo_d}) is of order $n^{-1}$ (terms of order $n^{-1/2}$ in
the r.h.s. of (\ref{lo_d}) are proportional to $\kappa_3$).

3) For Gaussian $W_{jk}$ the excess $\sigma$ is zero and (\ref{n2lo}) reduces
to  the asymptotic formula
\begin{eqnarray*}
{\bf E} \{g_n(z)\}=r(z)
\left[
1+ \frac{1}{n}
\frac{w^2r^2(z)}{[1-w^2r^2(z)]^2}
\right]
+O(n^{-2}).
\end{eqnarray*}
that has been derived earlier by the   formal diagrammatic
approach \cite{VWZ}.

%%%%%%%%%%%%%%%%%%%%%%%%%%%%% section 4 %%%%%%%%%%%%%%%%%%%%%%%%%%%%%

\section{Leading order of $F\lowercase{_n(z_1,z_2)}$}
\label{section4}

Let us recall our notation $F_n(z_1, z_2)$ (see (\ref{cov})) for the covariance
function of  $g_n(z)= n^{-1}\T (H-zI)^{-1}$,
\begin{eqnarray*}
F_n(z_1, z_2)={\bf E} \{g_n^C(z_1)g_n^C(z_2) \}={\bf E} \{g_n(z_1)g_n^C(z_2) \}
\end{eqnarray*}

In this section we prove the following

{\bf Theorem 2}
{\sl
Consider the Wigner ensemble of random  real
symmetric matrices with independent entries defined by  (\ref{H2W})-(\ref{W}).
Assume additionally that the
third and fourth moments of $W_{jk}$ do not depend on
$j$ and $k$ and that  $\widehat{\mu_5}=\sup_{j\le k}
{\bf E} \{|W_{jk}|^5\} < +\infty$

Let $f(z_1, z_2)$ be the function given by (\ref{f}).
If $z_1$ and $z_2$ belong to $U_0$ (\ref{U_0}), then the following  asymptotic
relation
\begin{eqnarray}
F_n(z_1, z_2) = n^{-2}f(z_1, z_2) + O(n^{-5/2})
\label{M}
\end{eqnarray}
is valid.
}

\vskip 1cm {\sl Proof.} Let us first prove (\ref{M}) under assumption
\begin{eqnarray}
\widehat{\mu_7} =\sup_{j\le k} {\bf E} \{|W_{jk}|^7\} < +\infty.
\label{7_th_mom}
\end{eqnarray}

Let $G_{jm}(z)$ denote a matrix element of $(H-zI)^{-1}$. By (\ref{res}),
\begin{eqnarray*}
z_1F_n(z_1, z_2)=\frac{1}{n} \sum_{j,m=1}^n
{\bf E} \{ H_{mj}G_{jm}(z_1)g_n^C(z_2)\}.
\end{eqnarray*}
For each pair $(j,m)$  $G_{jm}(z_1)g_n^C(z_2)$ is a smooth function of $H_{mj}$
 and its derivatives are bounded because of (\ref{Gn_b}). In particular,
$|D_{mj}^6 [G_{jm}(z_1) g^C_n(z_2)]| \le C (|\Im z_1|^{-1}+(|\Im z_2|^{-1})^8$.
Therefore by (\ref{im_nf}),
\begin{eqnarray}
\lefteqn{
zF_n(z_1,z_2)=
        } \label{E_M_n}\\
 & & \sum_{a=1}^5 \frac{1}{n^{(a+3)/2}}
\!\!\sum_{j,m=1}^n \!
\frac{\kappa_{a+1} (1+\delta_{jm})^{(a+1)/2} }{a!}
{\bf E} \{D_{mj}^a [G_{jm}(z_1)g_n^C(z_2)] \} + \varepsilon_n, \nonumber
\end{eqnarray}
where $\kappa_a$ are semi-invariants of $W_{mj}$, as in (\ref{next_g_n}), and
\begin{eqnarray*}
 |\varepsilon_n| \le n^{-5/2} C \hat \mu_7 (|\Im z_1|^{-1}+(|\Im z_2|^{-1})^8
\end{eqnarray*}

Performing differentiating in the r.h.s of (\ref{E_M_n}) one finds that the
sums of fifth, fourth and second  derivatives in the r.h.s. of (\ref{E_M_n})
contribute to $z_1F_n(z_1, z_2)$ terms of order $n^{-9/2}$, $n^{-7/2}$ and
$n^{-5/2}$, respectively (corresponding bounds can be obtained using
(\ref{var_g}), (\ref{Gn_b}) and (\ref{p_bound})).  So, these derivatives give
no contribution to the leading-order term of $F_n(z_1, z_2)$.  It remains to
estimate the contributions coming from first and third derivatives.

The contribution of third derivatives to $z_1F_n(z_1, z_2)$ consists of
several terms which we shall label by integer $a$ and $b$ satisfying $0 \le a,b
\le 3$ and $a+b=3$. These terms are
\begin{eqnarray*}
s^{(a,b)}_n(z_1,z_2) = \frac{\kappa_{4}}{6n^3}\sum_{j,m=1}^n
{\bf E}\{D_{mj}^a G_{jm}(z_1)D_{mj}^b g_n^C(z_2) \}.
\end{eqnarray*}

First estimate $s^{(3,0)}_n(z_1,z_2)$.
After differentiating it takes the form
\begin{eqnarray}
s^{(3,0)}_n(z_1,z_2)&=&-n^{-2}\kappa_4 {\bf E}
\Big\{
n^{-1}\sum_{j,m=1}^n G_{jm}^4(z_1)g_n^C(z_2) +6h_n(z_1)g_n^C(z_2)
\Big\}
\nonumber \\
 & & -n^{-1}\kappa_4 {\bf E} \{d^2_n(z_1)g_n^C(z_2) \}
\label{s_3_0}
\end{eqnarray}
($h_n(z)$ and $d_n(z)$ are defined in (\ref{fn}) and (\ref{cd}), respectively).
As it follows from (\ref{var_g}) and (\ref{p_bound}),  the mean value in the
r.h.s. of the equation above is $O(n^{-1})$, so
\begin{eqnarray*}
s^{(3,0)}_n(z_1,z_2)=
-n^{-1}\kappa_{4}{\bf E}\{d^2_n(z_1)g_n^C(z_2) \}+O(n^{-3}).
\end{eqnarray*}
Now we employ the obvious algebraic relation (in the below sub-index $C$
indicates subtracted mean value)
\begin{eqnarray}
{\bf E} \{\eta^2 \xi^C \}=2{\bf E} \{\eta^C \xi^C \}{\bf E} \{\eta \}
+{\bf E} \{(\eta^C)^2 \xi^C \}
\label{xieta}
\end{eqnarray}
and write ${\bf E}\{d^2_n(z_1)g_n^C(z_2) \}$ as
\begin{eqnarray*}
2{\bf E}\{d^C_n(z_1)g_n^C(z_2) \}{\bf E}\{d_n(z_1)\} +{\bf E}\{[d^C_n(z_1)]^2
g_n^C(z_2) \}.
\end{eqnarray*}
If $z \in U_0$, variances of $g_n(z)$ and $d_n(z)$ are of order $n^{-2}$. In
addition to this, $d_n(z)$ is bounded in absolute value by $C|\Im z|^{-2}$
for all realizations of $W_{jk}$. Therefore by the Cauchy-Schwarz inequality,
 ${\bf E}\{d^2_n(z_1)g_n^C(z_2) \}=O(n^{-2})$, provided  $z_1, z_2 \in U_0$.
Thus we have proved that $s^{(3,0)}_n(z_1,z_2)=
O(n^{-3})$. A similar argument shows that $s^{(0,3)}_n(z_1,z_2)$ and
$s^{(2,1)}_n(z_1,z_2)$ are $O(n^{-3})$, too.
The last term we need to estimate is
 $s^{(1,2)}_n(z_1,z_2)$. It is easy to see that
 \begin{eqnarray*}
\lefteqn{
s^{(1,2)}_n(z_1,z_2) = }\nonumber \\
 & & n^{-1}2\kappa_{4} {\bf E} \big\{
n^{-1}\sum_{m=1}^n G_{mm}(z_1)G_{mm}(z_2)
\big\}
{\bf E} \big\{
n^{-1}\sum_{j,m=1}^n G_{mm}(z_1)G_{jm}^2(z_2)
\big\} +O(n^{-3}).
\end{eqnarray*}
Mean values in the above are calculated in exactly the same way as
${\bf E} \{ c_n(z)\}$ and  ${\bf E} \{ d_n(z)\}$ have been done.
For large $n$:
\begin{eqnarray*}
n^{-1}2\kappa_{4} {\bf E} \big\{
n^{-1}\sum_{m=1}^n G_{mm}(z_1)G_{mm}(z_2) \big\}=
r(z_1)r(z_2) +O(n^{-1/2})
\end{eqnarray*}
and
\begin{eqnarray*}
{\bf E} \big\{
n^{-1}\sum_{j,m=1}^n G_{mm}(z_1)G_{jm}^2(z_2)
\big\}=
r(z_1)r^2(z_2) +O(n^{-1})
\end{eqnarray*}
(compare with (\ref{lo_c}) and (\ref{d_n_z})). Thus we conclude that the
contribution of third derivatives is
\begin{eqnarray}
-\frac{1}{n^2}\frac{\sigma r^2(z_1)r^3(z_2)}{1-w^2r^2(z_2)}
\label{lo_D_3}
\end{eqnarray}(\ref{nf})
(we recall using $\sigma $ for $\kappa_4$).

First derivatives in the r.h.s. of (\ref{E_M_n}) contribute to
$z_1F_n(z_1,z_2)$ the  term
\begin{eqnarray}
t_n(z_1,z_2)&=& -w^2{\bf E} \{g_n^2(z_1)g_n^C(z_2) \} -n^{-1}w^2
{\bf E} \{c_n(z_1)g_n^C(z_2) \} \nonumber \\
 & & -n^{-2}2w^2
{\bf E} \{n^{-1} \T (H-z_1I)^{-1} (H-z_2I)^{-2}\}.
\label{t_n}
\end{eqnarray}
By the resolvent identity (\ref{res_id})
\begin{eqnarray*}
(H-z_1I)^{-1}(H-z_2I)^{-1}=(z_1-z_2)^{-1}
\big[(H-z_1I)^{-1}-(H-z_2I)^{-1} \big]
\end{eqnarray*}
Thus one reduces ${\bf E} \{n^{-1}
\T (H-z_1I)^{-1} (H-z_2I)^{-2}\}$ to
\begin{eqnarray*}
(z_1-z_2)^{-1}\big[(z_1-z_2)^{-1}{\bf E} \{ g_n(z_1)-g_n(z_2)\}
- {\bf E} \{ c_n(z_2)\}\big].
\end{eqnarray*}
Now recalling (\ref{lo_m}), (\ref{lo_c}) and (\ref{quadra}),
\begin{eqnarray}
\lefteqn{
{\bf E} \{
n^{-1} \T (H-z_1I)^{-1} (H-z_2I)^{-2}
            \}
=
       } \nonumber \\
 & & \frac{1}{r(z_1)[1-w^2r^2(z_2)]}
\left[  \frac{r(z_1)-r(z_2)}{z_1-z_2} \right]^2 +O(n^{-1}).
\label{green_pr}
\end{eqnarray}

Clearly, ${\bf E} \{c_n(z_1)g_n^C(z_2) \} = {\bf E} \{c^C_n(z_1)g_n^C(z_2) \}$
and the corresponding summand in the r.h.s. of (\ref{t_n}) is $O(n^{-3})$. So
it remains to find ${\bf E} \{g_n^2(z_1)g_n^C(z_2) \}$.

Use (\ref{xieta}) to write
\begin{eqnarray}
{\bf E} \{g_n^2(z_1)g_n^C(z_2) \}&=& 2F_n(z_1,z_2)m^{(1)}_n(z_1) +
{\bf E} \{[g_n^C(z_1)]^2g_n^C(z_2) \} \nonumber \\
 &=& 2F_n(z_1,z_2)m^{(1)}_n(z_1) + O(n^{-5/2})
\label{triplet}
\end{eqnarray}
The latter equality uses ${\bf E} \{[g_n^C(z_1)]^2g_n^C(z_2) \}
=O(n^{-5/2})$, the bound which can be obtained following calculations of
Appendix \ref{appendixB}.

Now we are in a position to find the leading order of $F_n(z_1,z_2)$.
Collecting (\ref{lo_D_3})-(\ref{triplet}), we find that
\begin{eqnarray*}
z_1F_n(z_1,z_2)&=&-2w^2F_n(z_1,z_2)m^{(1)}_n(z_1)
-\frac{1}{n^2}\frac{\sigma r^2(z_1)r^3(z_2)}{1-w^2r^2(z_2)} \\
 & & - \frac{2w^2}{r(z_1)[1-w^2r^2(z_2)]}
\left[  \frac{r(z_1)-r(z_2)}{z_1-z_2} \right]^2 +O(n^{-5/2}).
\end{eqnarray*}
As it is clear from (\ref{lo_m}) and (\ref{quadra}),
\begin{eqnarray*}
[z_1+2w^2m^{(1)}_n(z_1)]^{-1}=-r(z_1)[1-w^2r^2(z_1)]^{-1}+O(n^{-1})
\end{eqnarray*}
and we end up with (\ref{M}).

The standard truncation technique of probability theory allows to prove
(\ref{M}) in the case when only 5th absolute moment of the random
variables $W_{jk}$ is uniformly bounded. Calculations using the truncation
technique are similar to those used in next section in proof of Theorem 3 and
we omit them. Theorem 2 proved.

One can consider the covariance function  $F_n(z_1, z_2)$ for the Wigner
ensemble of random Hermitian matrices (see remark 3 after the statement theorem
3 in next section). Repeating almost literally calculations used in proof of
theorem 2, one can prove that for the Wigner ensemble of Hermitian matrices
(\ref{M}) is still valid. The only difference is that now $f(z_1, z_2)$ is
given by the r.h.s. of Eq. (\ref{f}) multiplied  by factor $1/2$.

%%%%%%%%%%%%%%%%%%%%%%%%%%% sections 5   %%%%%%%%%%%%%%%%%%%%%%%%%%%%%

\section  {Gaussian fluctuations of the centralised trace of the resolvent}
\label{section5}

In this section we  prove the statement which is analogous to the central limit
theorem in the same sense in which the result (\ref{g2r}) is analogous to the
law of large numbers. Indeed, we can rewrite (\ref{g2r}) as following limiting
relation
\begin{eqnarray}
\li {1 \over n} \sum_{m=1}^n G_{mm}=r(z)
\label{lln}
\end{eqnarray}
valid with probability 1. Since the l.h.s. here has the form of the arithmetic
(Cesaro) mean, this relation is obviously similar to the
strong law of large numbers (or more generally to the  ergodic theorem). Common
wisdom of probability and ergodic theory
suggests that  (\ref{lln}) should imply that the probability distribution of
the random variable
\begin{eqnarray}
n^{1/2}
\Big[
n^{-1}\sum_{m=1}^n (G_{mm}-{\bf E}\{G_{mm})\}\Big] = n^{1/2}[g_n(z)-{\bf
E}\{g_n(z)\}]
\label{clt}
\end{eqnarray}
has the Gaussian form in the limit  $n=\infty $. We prove that under rather
natural conditions on $W_{jk}$ this is indeed the case provided that we use
non-standard normalisation, replacing $n^{1/2}$ in  (\ref{clt}) by $n$, i.~e.
we consider just the centralised trace of the resolvent
\begin{eqnarray}
\ga (z)=\sum_{m=1}^n (G_{mm}-{\bf E}\{G_{mm}\})=ng_n(z)- {\bf E}\{ng_n(z)\}
\label{ga}
\end{eqnarray}
instead of $n^{1/2}\ga (z)$. This normalisation can of course be anticipated
from the formula
(\ref{cov}) giving the order of magnitude (in fact, the asymptotics) of the
variance of  $g_n (z)$. This decay of the variance,
which is "twice" more strong than in the standard central limit theorem
setting, is rather typical for a number of
problems of the theory of disordered systems with non-local interaction and is
known as the strong self-averaging property
(see e.~g. Refs.\ \onlinecite{Be} and \onlinecite{P_MF} ).

\vskip 0.5cm

{\bf Theorem 3}
{\sl
Consider the Wigner ensemble of random  real symmetric
matrices with independent entries defined by  (\ref{H2W})-(\ref{W}) assuming
additionally that the fourth moments of $W_{jk}$ exist and
are independent of $j$ and $k$ and that the probability distribution functions
$ P_{jk}(w)$ of $W_{jk}$
satisfy the condition:
for any fixed fixed $ \nu  >0$
\begin{eqnarray}
\lim_{n \rightarrow \infty} \frac{1}{n^2}\sum_{j\le k}\; \int_{|x|>\nu n^{1/2}}
\;
x^2 d\mbox{Prob}[W_{jk} \le x] =0, \;\; \mbox{for any} \;\; \nu >0.
\label{(3.3)}
\end{eqnarray}
Then for any $z$ from $U_0 = \{z \in \C: \vert \Im z \vert \ge 2w\}$
the random function
$\ga (z)$ (\ref{ga})
converges in distribution as  $n\to \infty $  to the Gaussian random
function $\gamma (z)$ with zero mean and the covariance function  $f(z_1,z_2)$
given by (\ref{f}). In other words, for any integer $q$ and arbitrary
collection $z_1, \ldots  ,
z_q $ of complex numbers from $U_0$ the joint probability distribution of
random variables
$ \ga (z_1), \ldots  ,\ga (z_q)$ converges as $n\to\infty $ to the
$q$-dimensional
Gaussian
distribution with zero mean and the covariance matrix
$[f(z_s,z_t)]_{s,t=1}^q$

}

\vskip 1cm {\bf Remarks}

\vskip 0.5cm

1. Limit theorems concerning $\gamma^{(n)}(z)$  for the Wigner ensemble  were
established for the
first time by Girko (see Ref.\ \onlinecite{Girko} and references therein)
under assumption that there exist a positive $\delta$ such that
\begin{eqnarray}
\sup_{j\le k } \E |W_{jk}|^{4+\delta} <\infty ,
\label{(3.6)}
\end{eqnarray}
which is slightly more restrictive than (\ref{(3.3)}).  For example in the case
of identically distributed $W_{jk}$,
(\ref{(3.3)}) is obviously satisfied if $w_4\equiv {\bf E}\{W_{jk}^4\}$ is
finite. However, the more important in our opinion
improvement of
the result of  Ref.\ \onlinecite{Girko}  is that we calculate the covariance
matrix of the limiting
Gaussian process in the explicit form while in Ref.\ \onlinecite{Girko}
this matrix  was  given in the implicit form as a solution of a system of
cumbersome partial
differential
equations.

\vskip 0.5cm
2. For the random variables $W_{jk}$ satisfying (\ref{(3.6)}) we can estimate
the rate of convergence:
\begin{eqnarray}
\sup_{z_1,z_2 \in U_0} |\E\{\ga(z_1)\ga(z_2)\} - f(z_1,z_2)| =
O(n^{-\delta/2}).
\label{(3.7b)}
\end{eqnarray}

\vskip 0.5cm
3. Consider the Wigner ensemble of the $n \times n$ random Hermitian matrices
defined as in (\ref{H2W}) with
$W_{jk} =
A_{jk} +
i
B_{jk},
\;\; j\le k$,
$ W_{jk} = W_{kj}^\dagger$,
where $A_{jk}$ and $B_{jk}$ are mutually
independent random variables with zero mean, variance $w^2/2$ and excess
$\sigma /2$. It can be proved by analogous technique that in
this ensemble the fluctuations of the trace of the resolvent around its mean
become Gaussian in the limit $n \to \infty$.
The corresponding covariance function is given by (\ref{f}) in which   the
factor 2  is replaced by 1 in the denominator of both terms.

\vskip 1cm {\sl Proof.}
We shall work with real-valued variables $\a(z) =\Re \ga (z)$ and $\b(z) =\Im
\ga (z)$. Then we have to prove that the
limiting random functions $\alpha (z)$ and $\beta (z)$ are jointly Gaussian,
i.e. if
\begin{eqnarray*}
X(z,c)=
\left\{
\begin{array}{c c}
      \alpha (z) & \;\;\mbox{if} \;\; c= \alpha; \\
      \beta (z)  & \;\;\mbox{if} \;\; c=\beta ,
\end{array}
\right.
\end{eqnarray*}
and
\begin{eqnarray*}
(a(c),b(c))=
\left\{
\begin{array}{c c}
     (1/2,1/2) & \;\;\mbox{if} \;\;  c=\alpha; \\
    (1/2i,1/2i)& \;\;\mbox{if} \;\;  c=\beta ,
\end{array}
\right.
\end{eqnarray*}
then ${\bf E}\{X{(z,c)}\}=0$ and  for any integer $q$ and arbitrary collections
$z_s,s=1,...,
q, \,\,z_{s}\in U_0 $ and $c_s,,s=1,...,q, \,\, c_s\in\{\alpha,\beta \}$ the
joint probability distribution of random variables $X(z_1,c_1), \ldots
,X(z_q,c_q)$ is the $q$-dimensional Caussian distribution with
zero mean and covariance  matrix
\begin{eqnarray}
\E\{X (z_s,c_s)X(z_t,c_t)\}& =& a(c_s) a(c_t) f(z_s,z_t) + a(c_s) b(c_t)
f(z_s,z_t^\dagger)+ \nonumber \\
 & &
a(c_t) b(c_s) f(z_s^\dagger, z_t) + b(c_s) b(c_t) f(z_s^\dagger, z_t^\dagger),
\label{var-cov}
\end{eqnarray}
Let us consider the characteristic function of random variables
$X(z_1,c_1),..,X(z_q,c_q)$ which we shall write in the form
\begin{eqnarray*}
e_q^{(n)} (T_q,C_q,Z_q) =
\E\{ \prod_{s=1}^q \exp \{ i \tau_s [a(c_s) \ga (z_s) +  b(c_s) \ga
 (z_s^\dagger)] \},
%\label{(3.16)}
\end{eqnarray*}
where
$T_q = (\tau_1, \ldots, \tau_q),\,\, C_q = (c_1,\ldots, c_q),\,\, Z_q =
(z_1,\ldots, z_q)$

Recall that we designate the complex conjugate by the symbol $\dagger$. Also
writing the characteristic function we shall often omit indices indicating its
dependence on $n$ and  some other variables provided  there  will arise no
confusion.

Obviously
\begin{eqnarray*}
 {\partial \over \partial \tau_s }
\E \{ e_q (T_q)\} = i
\E \{e_q [a(c_s) \ga (z_s) +  b(c_s)  \ga (z_s^\dagger)]\}.
\end{eqnarray*}
Our aim is to show that there exists  a set of the
``covariance'' coefficients $A^{(n)}_{st}$, $s,t=1,...,q$ such that
for each fixed $T_q$
\begin{eqnarray*}
\li \left|\E \{e_q^{(n)}[a_s\ga (z_s)  +  b_s  \ga (z_s^\dagger)]\} -
i
\sum_{t=1}^{q} \tau_s A^{(n)}_{st}\E \{e_q^{(n)}\}\right| = 0,
\quad z \in U_0,
\end{eqnarray*}
to show that limits of all these  coefficients exist
\begin{eqnarray}
A_{st} = \li A_{st}^{(n)},
\label{AB}
\end{eqnarray}
and correspond to the r.h.s. of (\ref{var-cov}). Then standard arguments will
allow us to prove that the limit
characteristic function has the Gaussian form
$\exp(-1/2\sum_{s,t=1}^{q}  A_{st}\tau_s \tau_t)$.

Thus, we have   to compute
\begin{eqnarray*}
\E \{ e_q \ga (z) \} =
\sum_{j=1}^n \E\{e^C_{q}(Z_q) G_{jj}\},
\end{eqnarray*}
for large $n$ (we recall that sub-index $C$ indicate centering to zero mean).
Then putting one of $z_1, \ldots  ,z_q$ or one of their conjugates in place of
$z$ we calculate the limits in (\ref{AB})

We have complex energies $z, z_1, \ldots  ,z_q$ and  we introduce notation
$G(z_s)$ for the resolvent  corresponding to $z_s$ keeping notation
$G$ for the resolvent corresponding to $z$.

By the resolvent identity (\ref{res}),
\begin{eqnarray}
\sum_{j=1 }^n  \E  \{ e^C_q  G_{jj}  \} =
z^{-1}\sum_{j,m=1 }^n \E \{
e^C_q   G_{jm} H_{mj}  \}.
\label{(3.8)}
\end{eqnarray}
We compute the average in the r.h.s of (\ref{(3.8)}) following the scheme
described in
Section 2. However its direct application   requires
too strong conditions  on the distribution of $W_{jk}$.
 Thus, we modify slightly the general scheme  and carry out more accurate
estimates.

Denote by $\E_{mj}$ the conditional  expectation $\E \{ \cdot |W_{mj}=w\}$ and
rewrite the right-hand side of  (\ref{(3.8)}) in the form
\begin{eqnarray*}
z^{-1}  \n  \sum_{j,m=1 }^n \int \E_{mj} \{ e^C_q   G_{jm}\} w d
 P_{mj}(w).
\end{eqnarray*}
We split the integral
into the two ones over sets
$\Gamma_1 (n) = \{\omega: \, |W _{mj} |\le \delta n^{1/2}\}$  and
$\Gamma_2 (n) = \{\omega:\,  |W _{mj}|> \delta n^{1/2}\}$.
Now the
inequalities
\begin{eqnarray*}
 \left|\n \sum_{j,m=1}^n  \int_{\Gamma_2(n)} \E_{mj} \{ e^C_q  G_{jm}\} w
 P_{mj}(w) \right| &\le &
\eta^{-1} \n \sum_{j,m=1}^n \int_{\Gamma_2(n)}  |w|  d
 P_{mj}(w) \\
  & \le &
\eta^{-1}\nu^{-3} n^{-2} \sum_{j,m=1}^n \int_{\Gamma_2(n)}  w^4  d
 P_{mj}(w)
\end{eqnarray*}
and the assumption (\ref{(3.3)}) imply that only the integrals over
$\Gamma_1(n)$ gives the
a non-vanishing contribution contribution to (\ref{(3.8)}).

Following our general scheme, we expand the function
$e^C_q G_{jm}$ in powers of random variable
$\H_{mj} = \n W_{mj}$ restricted to $\Gamma_1(n)$. Since it is bounded in
absolute value by $\nu$  we can  write the relation
\begin{eqnarray}
 \n \sum_{j,m=1}^n \int_{\Gamma_1} \E_{mj}\{ e^C_q   G_{jm}\}
w d  P_{mj}(w)  =
 \sum_{k=1}^5 S_k(n),
\label{(3.9)}
\end{eqnarray}
 where
\begin{eqnarray*}
 S_k(n) = n^{k/2}\sum_{j,m=1}^n  \E
\left[e^C_q  G_{jm}\right]_{mj}^{(k-1)} \int_{\Gamma_1}   w^{k}
d P_{mj}(w),
\end{eqnarray*}
and $[...]_{mj}^{(k)}$ denotes that the $k$-th  derivatives with respect to
$\H_{mj}$ is taken and then $\H_{mj}$ is replaced by zero.
Let us note also that in
$S_5$ the expression in square brackets  is taken at some point $\tilde H_{mj}
\in (0,\nu)$.

The term $S_1(n)$ vanishes as $n\to\infty $ due our assumption (\ref{(3.3)}):
\begin{eqnarray*}
| S_1(n)|
&\le &
\left|\n \sum_{j,m=1}^n  \E  \left[ e^C_q  G_{jm}\right]_{mj}
\int_{\Gamma_1}   w
d P_{mj}(w)\right| \\
&\le &
\eta^{-1} \nu^{-3} n^{-2} \sum_{j,m=1}^n
\int_{\Gamma_1}   w^4 d P_{mj}(w)
%\label{(3.10)}
\end{eqnarray*}
where $\eta \equiv \vert \Im z\vert $.

The term $S_5(n)$ vanishes as $n \to \infty$ because it can be estimated by
\begin{eqnarray*}
 \frac{B_4(T_p)}{ \eta^{6} n^{5/2}} \sum_{j,m=1}^n \int_{\Gamma_1}   |w|^5
d P_{mj}(w) \le
 \frac{B_4(T_p) \nu}{ \eta^{6} n^{2}} \sum_{j,m=1} \int_{\Gamma_1}    w^4
d P_{mj}(w),
\end{eqnarray*}
where $B_4(T_p)$ is the upper bound of absolute value of the fourth derivative
in (\ref{(3.9)}).  For any fixed  $T_p$,  $B_4(T_p)$ is finite  and  recalling
(\ref{(3.3)}) we see that $S_5(n)$
goes to zero as $n \to \infty$

The term $S_3(n)$ also vanishes as $n\to\infty $. We  establish this fact at
the end of the proof.

Terms
$S_2(n)$ and
$S_4(n)$ give main contribution to (\ref{(3.8)}). Let us first consider
$S_2(n)$.
The resolvents $G(z_s)$ and $G$ are complex symmetric matrices and we have:
\begin{eqnarray}
- \sum_{j,m=1}^n \left[ e^C_q G_{jm} \right]_{mj}^{(1)} &=&
\left[  e^C_q G_{jj} G_{mm}
\right]_{mj} + \sum_{j,m=1}^n  \left[
e^C_q G^2_{jm}
     \right]_{mj} + \nonumber \\
 & &
i
\sum_{j,m=1}^n  \left[
\sum_{s=1}^q 2 \tau_s
     \left(
a_s [G^2(z_s)]_{jm} +b_s
[G^2(z_s^\dagger)]_{jm}
     \right)
G_{jm} e_q
\right]_{mj}.
\label{(3.11)}
\end{eqnarray}
Each term of the right-hand side of  this relation is a function of $\H$ in
which $\H_{mj}$ is replaced by zero and we have to come "back" to expressions
dependent
on  the whole matrix $\H$. To this end, we use again the resolvent identity but
now in the "opposite " direction.
We obtain for the first term  of  ({\ref{(3.11)})
\begin{eqnarray*}
 w^2\N \sum_{j,m=1}^n\E \{
\left[ e^C_q  G_{jj} G_{mm}
\right]_{mj} \} =
v^2\N \sum_{j,m=1}^n \E \{ e^C_q  G_{jj} G_{mm}\} -w^2\Ps,
\end{eqnarray*}
where
\begin{eqnarray*}
 \Ps  &=& n^{-3/2} \sum_{j,m=1}^n  \E  \{\left[  e^C_q  G_{jj}\,
G_{mm}
\right]_{mj}^{(1)}\}
\int_{\Gamma_1}   w
d P_{mj}(w) - \\
 & &  2^{-1} n^{ -2} \sum_{j,m=1}^n  \E \{ \left[ e^C_q  G_{jj}
\,G_{mm}\right]_{mj}^{(2)}\}
\int_{\Gamma_1}   w^2
d P_{mj}(w) - \\
 & &  6^{ -1} n^{-5/ 2} \sum_{j,m=1}^n  \E\{  \left[ e^C_q    G_{jj}\,
G_{mm}\right]^{(3)}
W_{mj}^3\}.
\end{eqnarray*}
It is easy to see that the first and the last terms of $\Ps$ vanish
as $n\to \infty $ due to our assumption (\ref{(3.3)}).

Using (\ref{(3.3)}) and  (\ref{D3}), we can rewrite the second term of $\Ps$
in the form
\begin{eqnarray*}
 \frac{-2i w^2}{  n^{2}}\sum_{j,m=1}^n \!\! \E \left[ e_q
 \!\!\sum_{s=1}^q
 \tau_s \left( a_s [G^2(z_s)]_{mm}[G(z_s)]_{jj} +b_s  [G^2(z_s^\dagger)]_{mm}
[G(z_s^\dagger)]_{jj}  \right) G_{jj} G_{mm}\right]_{mj} +
\P ,
\end{eqnarray*}
where the remainder $\P$ includes terms which have one or more factors $G_{jm}$
 or
terms of the form
\begin{eqnarray*}
n^{-2}\sum_{j,m=1}^n \E \{\left[e^C_q [G(z_s)_{jj}]^2 [G(z_j)_{jj}]^2
\right]_{mj}\}.
\end{eqnarray*}
It is clear that in all these expressions we can remove square brackets
$[...]_{mj}$ because this procedure will add terms of order $O(\n)$ to the
sums under consideration.  Using  the estimate   (\ref{p_bound}), taking into
account   the self-averaging
property
\begin{eqnarray}
\E |g_n^C(z)|^2 = o(\N),\,\, \hbox{as} \,\,n\to\infty
\label{(3.13a)}
\end{eqnarray}
(see Lemma 1 of Appendix \ref{appendixB} for the proof), and relation
\begin{eqnarray}
\li \E  |
(\N\sum_j G^\alpha_{jj}\, G^\beta_{jj}
\N \sum_j G^\mu_{mm} G^\nu_{mm})^C|  = 0
\label{(3.13b)}
\end{eqnarray}
with some $\alpha,\beta,\mu,\nu = 0,1,2$ (Lemma 2 of Appendix \ref{appendixB}),
it is easy to prove that
$\P $ also vanishes as $n\to \infty$.

We obtain finally that among terms coming from first summand in the right-hand
side of (\ref{(3.11)}) only the following
\begin{eqnarray}
\lefteqn{
w^{2}z^{-1}\E\{  e^C_q \sum_{j=1}^n G_{jj} \} g_n(z) -
} \label{(3.14)} \\
 & &
\frac{-2i w_4}{ n^{2}}
\E\{e_q\}  \sum_{s=1}^q \tau_s
\E\{\sum_{j,k=1}^n
 \left( a_s [G^2(z_s)]_{mm}[G(z_s)]_{jj} +b_s [G^2(z_s^\dagger)]_{mm}
[G(z_s^\dagger)]_{jj}  \right) G_{jj} G_{mm} \}.
\nonumber
\end{eqnarray}
does not vanish as $n \to \infty$.

The second summand in the r.h.s. of (\ref{(3.11)})
vanishes as $n\to\infty $. This becomes clear after
applying the same procedure of removing square brackets
$[\ldots]_{mj}$ to the  expression
$ n^{-2} \sum_{j,m=1}^n [e^C_q (G_{jm})^2 ]_{mj}^{(k)}
$ and using the estimates (\ref{p_bound}) and
(\ref{(3.13b)}).

Let us consider the contribution of the  last summand in  r.h.s of
(\ref{(3.11)}) for a fixed
parameter $z_s , s=1,...,q$.  Taking into account (\ref{(3.3)}) and repeating
the ``returning''
procedure, we obtain for this term
\begin{eqnarray*}
\lefteqn{
\N\sum_{j,m=1}^n \left[ [G^2(z_s)]_{jm} G_{mj} e_q\right]_{mj}
 = } \\
 & & \N\sum_{j,m=1}^n [G^2(z_s)]_{jm} G_{mj} e_q -
 n^{-3/2}\sum_{j,m=1}^n \left[ [G^2(z_s)]_{jm} G_{mj}  e_q\right]_{mj}^{(1)}
\int_{\Gamma_1}  w
d P_{mj}(w) - \\
 & &
n^{-2}\sum_{j,m=1}^n \left[ [G^2(z_s)]_{jm}  G_{mj} e_q\right]_{mj}^({2)}
\int_{\Gamma_1}   w^2
d  P_{mj}(w) -
n^{-5/2}\sum_{j,m=1}^n \E\{\left[ [G^2(z_s)]_{jm}  G_{mj}
e_q \right]^{(3]} W^3_{mj}\}.
\end{eqnarray*}

It is easy to see that in this equality terms with the first and the third
derivatives vanish as $n\to\infty $.  The second derivative gives
\begin{eqnarray*}
2\left[ \{
[G^2(z_s)]_{jj}\, G_{jj}\, G_{mm}\, G_{mm} + [G^2(z_s)]_{mm} \,G_{mm}\,
[G(z_s)]_{jj}\, G_{jj}
\} e_q
\right]_{mj}
\end{eqnarray*}
and 24 terms having a factor of the form $(G^\alpha)_{jm},\,\alpha =1,2$.
Omitting
brackets $[...]_{mj}$ and using (\ref{(3.13a)})-(\ref{(3.13b)}),  we see that
the last
term of (\ref{(3.11)})  gives
the leading contribution
\begin{eqnarray}
\lefteqn{
\E\{  e^C_q \} \sum_{s=1}^q \tau_s
\E\{ a_s \N \T G^2(z_s)G + b_s \N \T G^2(z_s^\dagger) G \}
        } \label{(3.15)} \\
 & &  -i w_4
\E\{  e^C_q \}
\sum_{s=1}^q \tau_s
\E\{ n^{-2} \sum_{j,m=1}^n (a_s [G^2(z_s)]_{jj}[G(z_s)]_{mm} +b_s
[G(z_s)]_{jj} [G^2(z_s)]_{mm})
\,G_{jj}  G_{mm}\}. \nonumber
\end{eqnarray}

Consider now the term $S_4(n)$ of (\ref{(3.9)}). The third derivative
$[e^C_q G_{jm}]'''_{mj}$ consists of 140 terms. One part of them
vanishes as $n\to\infty $ due to the property (\ref{(3.13a)})-(\ref{(3.13b)}),
another part - due to the
presence of the  factor of the form $(G^\alpha)_{jm},\,\alpha =1,2,3$. Only six
terms of
the form
\begin{eqnarray*}
i w_4
\sum_{s=1}^q
\E\{  e^C_q \} \E\{ n^{-2} \sum_{j,m=1}^n (a_s [G^2(z_s)]_{jj}[G(z_s)]_{mm}
+b_s
[G(z_s)]_{jj} [G(z_s)^2]_{mm})
\,G_{jj}  G_{mm}\}
\end{eqnarray*}
 are non-vanishing  in the limit $n\to\infty $. These terms arise  when we
differentiate $G_{jm}$ once and $e_q$ twice with respect to $\H_{mj}$.
Combining these terms with (\ref{(3.14)}) and (\ref{(3.15)}), we finally obtain
that
\begin{eqnarray}
\E \{e^C_q \T G  \} & = & i  {1\over z - 2w^{2}g_n(z)}
\E\{  e^C_q \}  \sum_{s=1}^q \tau_s ( 2w^2 \N\E \{\T  (a_s G(z_s)^2 +
b_s
G(z_s^\dagger) ^2)
\, G \} +  \label{(3.16)} \\
 & &
 2\sigma \E \{n^{-2}\sum_{j,m=1} (a_s G^2(z_s)_{jj}G(z_s)_{mm}+b_s
G^2(z_s^\dagger)_{jj}G(z_s^\dagger)_{mm})
 G_{jj} G_{mm}\} ).
\nonumber
\end{eqnarray}

Notice, that the denominator in the first term of the r.h.s. of this expression
is bounded away from zero because z belongs to the domain (\ref{U_0}).
Now, combining  (\ref{g2r}) and (\ref{r})
with relations
\begin{eqnarray}
\li \E \{\N \sum_{j=1}^m [G^2(z_s)]_{jj} G_{jj} \} =
r^2(z_s)[1-w^2r^2(z_s)]^{-1} r(z),
\label{(3.17)}
\end{eqnarray}
we derive from (\ref{(3.16)}) the final form of the covariance.

Relation (\ref{(3.17)})    can be easily deduced from our proof of
(\ref{(3.13a)})-(\ref{(3.13b)}).

Let us briefly discuss now the  proof of the fact
that
$S_3(n)$ of (\ref{(3.9)}) vanishes as
$n\to \infty $. The second derivative $[ e^C_q G_{jm}]_{mj}^{(2)}$  gives terms
each having the factor of the form
$[(G^\alpha)_{jj} (G^k)_{jm} (G^\beta)_{mm}]_{mj}$. The  brackets can be simply
omitted because the ``returning'' procedure adds terms of order $O(\n)$. Now,
regarding $\n (G^\alpha)_{jj}$ as vectors and
$(G^m)_{mj} $ as the kernel, we can write inequality
\begin{eqnarray*}
 |n^{-3/2} \sum_{j,m=1}^n (G^\alpha)_{jj} (G^m)_{mj} (G^\beta)_{mm}|\le
\eta^{-\alpha-\beta-m}\n
\end{eqnarray*}
which completes the proof of Theorem 3.

%%%%%%%%%%%%%%%%%%%%%%%%%%% sections 6   %%%%%%%%%%%%%%%%%%%%%%%%%%%%%

\section{Scaling limit and universality conjecture}

\label{section6}

We have presented above the  rigorous derivation of asymptotic corrections (in
fact expansions) for moments and
more complex quantities constructed from the traces of the
Green functions of the Wigner random matrix ensembles.
Now we use our result to draw certain non-rigorous conclusions
on the form of the leading term of the correlation function
$S_n(E_1,E_2)$ of the formal level density
$\rho_n (E)=n^{-1}
\mbox{tr} \;\delta (H-EI)$. Since $\rho_n(E ) = N'_{_n}(E)$ where $N_{n}(E)$ is
defined in (\ref{necf}), then basing on the relation (\ref{N2N_sc}) one can
conclude that the number of eigenvalues lying
inside the interval
$(E_1,E_2)$ with the center at $E$ will be
$N(E_2) - N(E_1) \sim n\rho_n(E) (E_2 - E_1)$, i.e. that the mean distance
between levels is $[n\rho_n(E)]^{-1}$. Thus the scaling $E_2-E_1 = O(n^{-1})$
defines the microscopic or local regime in which one deals with a finite
numbers of eigenvalues \cite{Mehta,Dy}.

Consider the density-density correlation function
\begin{eqnarray}
S_{n}(E_1,E_2)  = \E \{\rho_n(E_1) \rho_n(E_2)\} -
\E \{\rho_n(E_1)\}\E \{\rho_n (E_2)\}.
\label{(4.1)}
\end{eqnarray}
By using (8) and (9) we obtain from (14) and (\ref{ga}) that the Stieltjes
transform of $S_n(E_1,E_2)$
\begin{eqnarray*}
 F_n(z_1,z_2) = \int\!\int {S_n(\l_1,\l_2) \over (\l_1-z_1)(\l_2-z_2)}
d\l_1 d\l_2,
\quad \Im z_i \neq 0
%\label{F_n}
\end{eqnarray*}
is
\begin{eqnarray*}
F_n(z_1,z_2) = n^{-2}\E \{\ga(z_1) \ga(z_2)\}.
%\label{F_n1*}
\end{eqnarray*}
It follows from the inversion formula for the  Stieltjes transform
$f(z) = \int (\l-z)^{-1} \rho (\l )d\l$
\begin{eqnarray}
\rho(E) =
\pi^{-1} \lim_{\epsilon \downarrow 0}
\Im f(E + i \epsilon)  \equiv  I_{E_1}\{f(z)\}
\label{FB1}
\end{eqnarray}
that to find $S_n(\l_1,\l_2)$ ,  one has to know $F_n(z_1,z_2)$ up to the real
axis in both variables because
\begin{eqnarray}
S_n(E_1,E_2)= I_{E_1}\circ I_{E_2}\{F_n(z_1,z_2)\}
\label{FB2}
\end{eqnarray}
On the other hand, we have found the form (\ref{cov}) and (\ref{f}) of
$F_n(z_1,z_2)$ only in the domain  $\vert \Im z| \ge 2w$. However, since the
function $f(z_1,z_2)$ given by (\ref{f}) can
obviously be continued up to the real axis with respect to the both variables
$z_1$
and $z_2$ we can apply to the first term of (\ref{cov}) the operation
$I_{E_1}I_{E_2}$, $E_1 \ne E_2$ to compute formally the ``leading'' term of
the density-density correlation function. This means that we
perform
first the limit $n \to \infty$ and then the limits $\epsilon_1, \epsilon_2
\downarrow 0$. This order of limiting transitions is inverse with respect
to that prescribed by the definition of this correlation function.

To make these computations , we use the identity
\begin{eqnarray*}
{r_1 - r_2\over z_1-z_2} = {r_1 r_2\over 1-w^2r_1r_2}
\end{eqnarray*}
which follows from  (\ref{r}) or (\ref{quadra}). The identity yields the
relations $
\ve |r(\l+i\ve)|^2 = \Im r(\l+i\ve) (1-w^2 |r(\l+i\ve)|^2)
$
and
$
 |r(\l+i\ 0)|^2= w^{-2}$ for $E$  such that $ \Im r(\l+i
0) >0$.
Combining these relations with (\ref{rho_semi}), we obtain that
\begin{eqnarray*}
w^2[\Re\, r(\l+i 0)]^2 = {\l^2\over 4w^2}\,\,\,  \hbox{and}
\,\,\,w^2 [\Im r(\l+i 0) ]^2 = 1 -{\l^2\over 4w^2}.
\end{eqnarray*}

Using these equalities, we derive from our result (\ref{cov}) and (\ref{f})
and from  (\ref{FB1}) and (\ref{FB2})that
\begin{eqnarray}
S_n(E_1,E_2)&=&
-\frac {1}{\beta \pi^2 [n(\l_1-\l_2)]^2}
\frac {4w^2 - \l_1\l_2}{(4w^2-\l_1^2)^{1/2}(4w^2-\l_2^2)^{1/2}}  \nonumber \\
 & & + \frac {\sigma}{2n^2 \pi^2w^8}\frac {(2w^2-\l_1^2)(2w^2-\l_2^2)}
{(4w^2-\l_1^2)^{1/2}(4w^2-\l_2^2)^{1/2}}.
\label{(4.8)}
\end{eqnarray}
with $\beta=1$. It can be shown that  for the Hermitian matrices with
independent entries (see Remark 3 to Theorem 3) the density-density correlator
has the same form with $\beta =2$. For the Gaussian orthogonal and unitary
ensembles (GOE and GUE) $\sigma  =
0$, and we recover the result
\begin{eqnarray*}
S_n(E_1,E_2) =
 -\frac {1}{\beta \pi^2 [n(\l_1-\l_2)]^2}
\frac {4w^2 - \l_1\l_2}{(4w^2-\l_1^2)^{1/2}(4w^2-\l_2^2)^{1/2}}
\end{eqnarray*}
obtained in Ref.\ \onlinecite{FMP} and Ref.\ \onlinecite{Pandey}.

We see that in a general
non-Gaussian case the respective expression depends not only on the second
moment of entries , but also on their  fourth moment via the excess $\sigma$.

The remarkable fact is that this dependence vanishes  in the microscopic
(called also scaling) limit
\begin{eqnarray}
\l_1, \l_2 \to \l,\,\,n(\l_2-\l_1) \to s
\label{(4.4)}
\end{eqnarray}
Indeed, it easy to see that in this limit we obtain from
(\ref{(4.8)}) very simple expression:
\begin{eqnarray}
\lim_{n(\l_2-\l_1)\to s} S_n (\l_1,\l_2) = -{1\over \beta \pi^2 s^2}.
\label{(4.10)}
\end{eqnarray}
According to Wigner and Dyson (see e.g. Ref.\ \onlinecite{Mehta}), the exact
large-$s$
asymptotics for the limiting correlation function of the Gaussian
ensembles are:
$-1/(\pi^2s^2)$ (GOE) and
$-\sin^2\pi \rho(\l) s/(\pi^2s^2)$ (GUE). Comparing these  expressions with
our results, we see that the  procedure of computing of  the correlation
function yields for the general case the expression coinciding with the
large-$s$ asymptotics of the Gaussian  ensembles correlation function
smoothed over energy intervals  whose length is much smaller than the
macroscopic scale $w=\E\{W^2\}^{1/2} $ but much bigger than the microscopic
scale given by the mean level spacing $[n\rho(E)]^{-1}$. It is natural
to think that in our
computations the smoothing has been implemented ``automatically'' due to the
nonzero imaginary part of the spectral parameter $\Im z_j$. We notice that the
same procedure is widely used in the mesocsopic calculations based on the Kubo
formula, weak disorder perturbation theory, etc.

The  independence of the scaling limit  expressions (\ref{(4.10)}) on the
excess $\sigma$ can be regarded as a  support of
the universality conjecture for the Wigner ensembles. Let us mention
supports of this conjecture for other ensembles.

The first one  \cite{MF_2}
concerns the so-called sparse (or diluted) random matrices whose entries are
independently distributed random variables such that $\mbox{Pr}\,
\{H_{k,l}=0\} = p/n$. The authors of Ref.\ \onlinecite{MF_2} used the Grassman
integral technique and
found the Wigner-Dyson universal form of the density-density correlator if $p$
is large
enough.

The second \cite{PKV} concerns the ensemble
$H = \sum_{\mu =1}^p \, \tau_{\mu}(\cdot,\xi^{\mu})\xi^{\mu}$, where
$\tau_{\mu}$ and $\xi^{\mu}=\{\xi^{\mu}_1, \ldots  , \xi^{\mu}_n\}$ are
independent identically distributed random variables (the ensemble was
introduced in Ref.\ \onlinecite{MP}). For this ensemble, whose entries are
dependent random variables, the  analogue of (\ref{(4.8)}) is obtained
and it is shown  that its scaling limit is the  same as above.

The third follows from Appendix \ref{appendixA} below.
We consider there case of the deformed GOE (see definitions below). For
this ensemble the analog of Theorem 3  was proved in Ref.\ \onlinecite{K}. In
the appendix we present a short derivation of
the density-density correlator and
show that in the scaling limit it has the form (\ref{(4.10)}).

We mention also that for the unitary invariant ensembles of the form
(\ref{P_n(H)}) the universality conjecture is rigorously proved
in Ref.\  \onlinecite{PS} for a rather broad class of functions $F(x)$.

%%%%%%%%%%%%%%%%%%%%%%%%%%%%%%%%%%%%%%%%%%%%%%%%%%%%%%%%%%%%%%%%%%%%%%%

\section*{Acknowledgments}

This work was supported in part by the International Science Foundation under
Grant No U2S000.

Part of this work was done when one of the authors, BAK was at
Ruhr-Universit\"at-Bochum and he acknowledges financial support of Deutsche
Forschungsgemeinschaft under Grant No SFB237

\appendix

%%%%%%%%%%%%%%%%%%%%%%% Appendix A %%%%%%%%%%%%%%%%%%%%%%%%%%%%%

\section {Scaling limit for the deformed GOE}
\label{appendixA}

In this Appendix  we find the exact form of the leading term  of the covariance
function
$F_n(z_1,z_2)$ (\ref{cov}) for the ensemble
$
H_d = H^{(0)} + H$, where $H^{(0)}$ are $n\times n$ nonrandom matrices such
that there exists the "unperturbed" IDS
\begin{eqnarray*}
N^{(0)} (E ) =\li n^{-1}\#\{e_j:  e_j \;\; \mbox{is an eigenvalue of} \;
H^{(0)} \; \mbox{and } \; e_j \le E\}
\end{eqnarray*}
and $H$ belongs to the GOE ((\ref{P_n(H)}) with  $F= x^2/4w^2$). This ensemble
is called \cite{Brody}  the deformed GOE.
Because of the orthogonal invariance of the GOE distribution, we can restrict
our considerations to the case of diagonal $H^{(0)}$.
So we assume that $H^{(0)}=[\delta_{jk}e_j]_{j,k=1}^n $
and real numbers $e_j$ are such that the limit
\begin{eqnarray}
g^{(0)}(z)=\li \frac{1}{n} \sum_{j=1}^n \frac{1}{e_j-z}
\label{cp}
\end{eqnarray}
exists for all non-real $z$. The function $g^{(0)}(z)$ is  the  Stieltjes
transform of $N^{(0)}(E)$. We shall use
notation $d_j(z)$ for $(e_j-z)^{-1}$ and $g_n(z)$ for the normalized trace of
the resolvent of $H_d$.

Subsequent arguments are quite similar to those used in derivation of
 (\ref{var_g}) and (\ref{id_1}) - (\ref{gb}) for the GOE.
By using (17) and  (19) for $H_1=H^{(0)}$ and $H_2=H_d$ one can derive  the
following two relations ( analogues of
(\ref{id_1})  and  (\ref{id_2}) ) :
\begin{eqnarray}
{\bf E} \{ G _{jk}(z) \}= d_{j}(z)\delta_{jk} +
{\bf E} \{g_n(z) G_{jk}(z)\} d_{k}(z)+
\N\sum_{m} {\bf E} \{G_{jm}(z) G_{km}(z)\} d_{k}(z),
\label{(A.1)}
\end{eqnarray}
and
\begin{eqnarray}
{\bf E}  \{g_n^C (z_1) G_{jj}(z_2)\} &=&  w^2{\bf E}\{  g^C_n (z_1) G_{jj}(z_2)
\} {\bf E} \{ g_n (z_2)\}
d_{j}(z_2)+ \nonumber \\
 & &
w^2{\bf E} \{ g^C_n(z_1)g_n^C(z_2) G_{jj}(z_2)\} d_{j}(z_2)+ \nonumber \\
 & & w^2\N{\bf E} \{
g^C_n(z_1) [G^2(z_2)]_{jj} \} d_{j}(z_2) + \nonumber \\
 & &
2w^2n^{-2} \sum_m {\bf E} \{ [G^2(z_1)]_{jm} G _{mj}(z_2)\} d_{j}(z_2).
\label{(A.2)}
\end{eqnarray}

It follows from (\ref{(A.1)}) that if  for $z\in U_0$ where $U_0$
is defined in (\ref{U_0})
\begin{eqnarray}
\li {\bf E} \big\{ | g_n(z) - {\bf E} \{ g_n(z) \}| \big\}= 0,
\label{(A.3)}
\end{eqnarray}
then
$
\li {\bf E} \{ g_n(z)\} = g(z),
$ where $ g(z)$ is the unique solution of the functional equation \cite{P_TMP}
\begin{eqnarray}
g(z) = g^{(0)}( z+ w^2 g(z))
\label{(MF)}
\end{eqnarray}
satisfying $\Im g(z) \Im z \ge 0$. In the equation above, $ g^{(0)}(z)$ is
given by (\ref{cp}).

It is easy to show that
(\ref{(A.1)}) and (\ref{(A.3)}) imply the relation
\begin{eqnarray}
\sup_{j=1,\dots,n}|{\bf E} \{ G_{jj}(z) \}-  g^{(n)}_{j}(z)| = O(\N), \quad
z\in U_0,
\label{(A.4)}
\end{eqnarray}
where $g^{(n)}_{j}(z) $ solves  the equations
\begin{eqnarray}
g^{(n)}_{j}(z) = {1\over e{j} -z - w^2 g^{(n)}(z)},\,\, j=1, \ldots , n \quad
 g^{(n)}(z) = \N \sum_{m=1}^n g^{(n)}_{m}(z).
\label{MF_1}
\end{eqnarray}
Indeed, if  $ V^{(n)}_{j}={\bf E} \{ g^C_n (z_1) G_{jj}(z_2) \}$ then by
(\ref{(A.2)}),
\begin{eqnarray}
V^{(n)}_{j} &=&  w^2 V^{(n)}_{j} {\bf E} \{ g_n (z_2) d_{j}(z_2)\}+ \nonumber
\\
 & &
\N \sum_j V^{(n)}_j  {\bf E} \{ g_n (z_2) d _{j}(z_2)\}+ 2w^2n^{-2} {\bf E} \{
\sum_j
[G^2(z_2)]_{jm} G_{mj}(z_2) d_{j}(z_2)\}+ \nonumber \\
 & & w^2\N{\bf E} \{ g_n^C(z_1) [G^2(z_2)]_{jj} \} d_{j}(z_2) + w^2{\bf E} \{
g_n^C(z_1) g_n^C(z_2) [G_{jj}(z_2)]^C \} d_{j}(z_2).
\label{(A.5)}
\end{eqnarray}
Now, repeating arguments used at the end of Section II, we can easily obtain
the estimate
$
\N \sum_j V^{(n)}_{j} = O(n^{-2})
$ which proves (\ref{(A.4)}). Using this estimate and considering (\ref{(A.5)})
once more, we
obtain the estimate
\begin{eqnarray}
\sup_j |V^{(n)}_{j}| = O(n^{-2}).
\label{(A.6)}
\end{eqnarray}

It follows from the resolvent identity (\ref{res_id})  that
$$
\sum_j [G^2(z_1)]_{jm} [G(z_2)]_{mj} = {[G^2(z_1)]_{jj}\over z_1-z_2} - {
[G(z_1)]_{jj} - [G(z_2)]_{jj}\over (z_1-z_2)^2}.
$$
Taking into account this relation, (\ref{(A.4)}), and (\ref{(A.6)}), we obtain
that if
\begin{eqnarray*}
f_d (z_1,z_2) = \li n^2{\bf E} \{g^C(z_1) g^C(z_2) \},
\end{eqnarray*}
then
\begin{eqnarray*}
|\N \sum_j V^{(n)}_m - f_d(z_1,z_2) | = o(n^{-2}),\quad z\in U_0,
\end{eqnarray*}
and as the result
\begin{eqnarray}
f_d(z_1,z_2)& = & - {2w^2\over (z_1 - z_2)^2}
\li {\N \sum_m [g_m^{(n)}(z_1) -g_m^{(n)}(z_2)]g_m^{(n)}(z_2)\over  1-
w^2\N\sum_j
g_m^{(n)}(z_2)^2} \nonumber \\
 & & +
{1\over z_1-z_2} \li {\N \sum_j [G^2(z_1)](z_1) g_m^{(n)}(z_2)\over  (1-
w^2\N\sum_j  g_m^{(n)}(z_1)^2)(1- w^2 \N\sum_j g_m^{(n)}(z_2)^2)}.
\label{(A.7)}
\end{eqnarray}

Since
$\li \N\sum_m  g_m^{(n)}(z)^2 = \int (E - z - w^2  g_0(z))^{-2} dN_0(E)
\equiv \Phi_2$, then the second fraction of the last term of (\ref{(A.7)}) is
not
singular  for $z_i = E \pm i 0$ with $E$ such that $ \Im g(E+i 0) >0$. Thus,
this term vanishes in the scaling limit (\ref{(4.4)}).

Consider now the first term of the right-hand side of (\ref{(A.7)}). Simple
computation shows that
\begin{eqnarray*}
\N \sum_m g_m^{(n)}(z_1) g_m^{(n)}(z_2) =  { g^{(n)}(z_1) - g^{(n)}(z_2) \over
z_1-z_2 + w^2[
g^{(n)} (z_1) - g^{(n)}(z_2)]}.
\end{eqnarray*}
Since, according to (\ref{(A.3)}) - (\ref{MF_1})  $\li g^{(n)}(z) = g(z)$ we
find  for this term
\begin{eqnarray*}
-{2w^2\over (z_1-z_2)^2}
\left( {g_0(z_1) - g_0(z_2)\over z_1-z_2 +w^2[g_0(z_1) - g_0(z_2)]} -
\Phi_2\right) (1-w^2
\Phi_2)^{-1} + O(|z_1-z_2|^{-1}).
\end{eqnarray*}
 This relation  implies that in the scaling limit (\ref{(4.4)})  we obtain
again the simple universal expression
(\ref{(4.10)}) .

%%%%%%%%%%%%%%%%%%%%%%%% Appendix B %%%%%%%%%%%%%%%%%%%%%%%%%%%%%%%

\section{Auxiliary facts}
\label{appendixB}

{\bf Lemma 1. } {\sl Self-averageness property (\ref{(3.13a)}) holds under
assumptions of
Theorem 3}
\vskip 0.5cm
{\sl Proof. } We denote
\begin{eqnarray}
F_n(z,z^\prime)={\bf E} \{g^C (g')^C \} \equiv {\bf E} \{g^C g^\prime \}
\label{(B.1)}
\end{eqnarray}
where $g\equiv g_n(z)$ and $g^\prime\equiv g_n(z^\prime),\ g_n(z)=n^{-1}
\T (H-zI)^{-1}$ and $g^C=g-{\bf E} \{ g\}$.

Obviously, $
G_{jj}(z^\dagger)=G^\dagger_{jj}(z)
$ and $
F_n(z,z^\dagger)={\bf E} \{|g^C(z)|^2 \}$.

Let us apply the resolvent identity (\ref{res}) to the last factor $g^\prime$
in the right-hand side of (\ref{(B.1)}). We obtain the
relation
\begin{eqnarray}
F_n(z,z^\prime)={ 1\over z^\prime n} \sum_{jm}{\bf E} \{ g^C G^\prime_{jm}
\H_{mj} \},
\label{(B.2)}
\end{eqnarray}
where $G^\prime \equiv G(z^\prime)$. Comparing relations (\ref{(B.2)}) and
(\ref{(3.8)}), we
see that their right-hand sides are similar. The only difference is that
the sum in (\ref{(B.2)}) has extra factor $ n^{-1}$ and $e_q$ of (\ref{(3.8)})
is replaced by $g$.  Hence, one can
compute the average in (\ref{(B.2)}) in the same way
as it
was done for the right-hand side of (\ref{(3.8)}) and come  to the expression
$
F_n(z,z^\prime)=\sum_{k=1}^{5} T_k(n)
$
where $T_k(n)$ are similar to $S_k(n),\ k=1,..,5$ in (\ref{(3.9)}).

Thus we find that
$T_1(n)$ and $T_3(n)$ are of order $o(n^{-1})$, as $n\to\infty$  just as in the
case of $S_1(n)$ and $S_3(n)$.

Consider $T_5(n)$ which is analogous to $S_5(n)$ in (\ref{(3.9)}).
It contains four
derivatives of $G_{kk}G_{jm}$ by $H_{mj}$. It follows from (\ref{D3})
that
the result of differentiating includes at least one factor $G_{jm}$.
Combining (\ref{p_bound}) with inequalities used to estimate $S_5(n)$, we
easily derive that $T_5(n)$ is a quantity of order $O(n^{-3/2})$.

Let us estimate $T_4(n)$ acting in the same way as in the case of $S_4(n)$. As
it was mentioned in \ref{section5},
the non-vanishing contribution to $S_4(n)$ comes from terms arising from one
derivative of $G^\prime_{jm}$ and two derivatives of $e_q$. The rest of the
terms
are of order $o(1)$. Thus, in the corresponding terms of
$T_4(n)$ we have to take into account only terms with factors $G$ or $G^\prime$
having coincident arguments. It is easy to see that due to extra factor
$n^{-1}$ in front of the whole sum and factor $n^{-1}$ in $g(z)$, these
terms are of order $n^{-2}$. Thus, $T_4(n)$ is of order $o( n^{-1})$.

Turning to $T_2(n)$ and taking into account previous arguments, we arrive at
the relation
\begin{eqnarray}
F_n(z,z^\prime)=- {{w^2}\over{z^\prime n^2}}\sum_{j,m}{\bf E} \{  g^C
G^\prime_{jj}
G^\prime_{mm} \}
-{{w^2}\over{z^\prime n^2}}\sum_{j,m}{\bf E} \{ g^C G^\prime_{jm}G^\prime_{jm}
\}+
\Phi^\prime(z,z^\prime),
\label{(B.3)}
\end{eqnarray}
where $\Phi^\prime(z,z^\prime)=o(n^{-1})$. Using (\ref{p_bound}), we easily
obtain that
\begin{eqnarray*}
{{w^2}\over{n^2}}\left|\sum_{j,m}{\bf E} \{ g^C G^\prime_{jm}
G^\prime_{jm}\}\right|
&\leq &
{{w^2}\over{n^2}}\sum_j {\bf E} \Big\{|g^C| \sum_m | G^\prime_{jm}|^2  \Big\}
\\
 & \leq &
{{w^2}\over{n|\Im z^\prime|^2}} {\bf E}^{1/2} \{ | g^C|^2 \}
\end{eqnarray*}
Observing that
\begin{eqnarray*}
{\bf E} \{g^C g^\prime g^\prime \}=2{\bf E} \{g^C g^\prime \}
{\bf E}  \{g^\prime \} +{\bf E} \{ g^C
[g^\prime ]^C g^\prime \},
\end{eqnarray*}
we derive from (\ref{(B.3)}) that for $z^\prime=z^\dagger$ and $z \in U_0$
(\ref{U_0}):
\begin{eqnarray*}
C{_1}F_n(z, z^\dagger ) -  C_2n^{-1} |F_n(z,z^\dagger ) |^{1/2} -
|{\Phi^\prime}_n| \leq O,
\end{eqnarray*}
where $C_1$ and $C_2$ are absolute constants (cf.(\ref{qe}). This inequality
implies (\ref{(3.13a)}). Lemma  proved

\vskip 0.5cm

{\bf Lemma 2. } {\sl Under assumptions of Theorem 3 the relation (\ref{(3.13b)}
is true.}
\vskip 0.5cm
{\sl Proof. } It suffices to show that
\begin{eqnarray}
R_n\equiv{\bf E} \{ |G_2^C|^2 \} = o(1),\ n\to\infty,
\label{evm}
\end{eqnarray}
where $G_2\equiv n^{-1}\sum_j G^2_{jj}$. Repeating computations of
previous proof, we obtain the following relation
$$
R_n={{1}\over {z^\prime n} } \sum_{j,m}{\bf E} \{ G^C_2 G^\prime_{jj}
G^\prime_{jm}
H_{mj} \}.
$$
Comparing again the right-hand side of this equality with those of
(\ref{(B.2)})  and
(\ref{(3.8)}) and repeating the corresponding computation, we conclude that
\begin{eqnarray}
R_n=
- {w^2 \over  z^\prime n^2 }\sum_{j,m}{\bf E} \{ G_2^C
G^\prime_{jj}G^\prime_{jj}
G^\prime_{mm}\} + {{w^2}\over{z^\prime n^2}}\sum_{j,m}{\bf E} \{ G_2^C
G^\prime_{jj}
G^\prime_{jm}G^\prime_{mm} \} + {\Phi^{\prime\prime}}_n,
\label{(B.4)}
\end{eqnarray}
where ${\Phi^{\prime\prime}}_n=o(1)$ as $n \to\infty$ for $z^\prime=z^\dagger $
and
$|\Im z| > 0$. Taking into account (\ref{(3.13a)}), we derive from
(\ref{(B.4)})
that for $z\in U_0$
$$
C_3R_n - C_3 R_n -n^{-1} C_4
|R_n|^{1/2} + o(1) \leq 0,
$$
where $C_3$ and $C_4$ are some absolute constants (cf.(\ref{qe})).  This
inequality implies (\ref{evm}).
Lemma is proved.

%%%%%%%%%%%%%%%%%%%%%%%%%% references %%%%%%%%%%%%%%%%%%%%%%%%%%%%%


\begin{references}
\bibitem{ftnote} In fact, in Ref.\ \onlinecite{P_TMP} a somewhat different and
stronger form of condition (\ref{Lin}) was used.
\end{references}
\end{document}